\documentclass[preprint,showpacs,showkeys]{revtex4}

\usepackage{amsmath,amsfonts,amssymb,amsthm}
\usepackage{epsfig}    
\usepackage{graphicx}
\usepackage[ansinew]{inputenc}
\usepackage{times}

\def\a{\alpha}
\def\b{\beta}
\def\g{\gamma}
\def\d{\delta}
\def\ve{\varepsilon}
\def\sq2{\sqrt{\frac{\varepsilon_0}{\mu_0}} }  

\begin{document}

\title{Relativistic nature of a magneto\-electric modulus of
  Cr$_2$O$_3$-crystals: a new 4-dimensional pseudoscalar and its
  measurement}

\author{Friedrich W. Hehl}
\email{hehl@thp.uni-koeln.de} 
\affiliation{Institute for Theoretical Physics, University of Cologne,
50923 K\"oln, Germany}
\affiliation{Department of Physics and Astronomy,
University of Missouri-Columbia, Columbia, MO 65211, USA}
\author{Yuri N. Obukhov}
\email{yo@thp.uni-koeln.de}
\affiliation{Institute for Theoretical Physics, University of Cologne,
50923 K\"oln, Germany}
\affiliation{Department of Theoretical Physics, Moscow State University,
117234 Moscow, Russia}
\author{Jean-Pierre Rivera}
\email{Jean-Pierre.Rivera@chiam.unige.ch} 
\affiliation{Department of Inorganic, Analytical and
  Applied Chemistry,\\ University of Geneva, Sciences II, 30 quai E.
  Ansermet, CH-1211 Geneva 4, Switzerland}
\author{Hans Schmid}
\email{Hans.Schmid@chiam.unige.ch}
\affiliation{Department of Inorganic, Analytical and
  Applied Chemistry,\\ University of Geneva, Sciences II, 30 quai E.
  Ansermet, CH-1211 Geneva 4, Switzerland}

\begin{abstract}
  Earlier, the magnetoelectric effect of chromium sesquioxide
  Cr$_2$O$_3$ has been determined experimentally as a function of
  temperature. One measures the electric field-induced magnetization
  on Cr$_2$O$_3$ crystals or the magnetic field-induced polarization.
  From the magnetoelectric moduli of Cr$_2$O$_3$ we extract a
  4-dimensional relativistic invariant pseudoscalar $\widetilde{\a}$.
  It is temperature dependent and of the order of $\sim 10^{-4}/Z_0$,
  with $Z_0$ as vacuum impedance. We show that the new pseudoscalar is
  odd under parity transformation and odd under time inversion.
  Moreover, $\widetilde{\a}$ is for Cr$_2$O$_3$ what Tellegen's {\it
    gyrator} is for two port theory, the {\it axion} field for axion
  electrodynamics, and the PEMC (perfect electromagnetic conductor)
  for electrical engineering.
\end{abstract}

\maketitle
                          
\section{Introduction}\label{sec1}

Usually, in vacuum, the constitutive relations of classical
electrodynamics are $\mathbf{D}=\varepsilon_0\mathbf{E}$ and
$\mathbf{H}=\mathbf{B}/\mu_0$. The electric constant $\ve_0$
(permittivity of free space) alone has no direct meaning in
4-dimensional spacetime; the analogous is true for the magnetic
constant $\mu_0$ (permeability of free space). However, if we combine
both constants, the situation changes. As shown by Post \cite{Post},
for example, it is rather the square root of the quotient of both
constants, namely $Y_0:=\sqrt{\varepsilon_0/\mu_0}$, the {\it vacuum
  admittance} of $Y_0\approx1/377\, \Omega$ that represents a {\it
  scalar} in 4 dimensional spacetime in {\it arbitrary} coordinates;
the same is true for its reciprocal, the vacuum impedance (resistance)
$Z_0:=1/Y_0$.  Thus, it is possible to extract 4-dimensional
information from both 3-dimensional constants, provided they are taken
together.

Moreover, $c:=1/\sqrt{\ve_0\mu_0}$, the vacuum speed of light, has
also a 4-dimensio\-nal meaning, even though $c$ is only a scalar under
Poincar\'e (inhomogeneous Lorentz) transformations. This is obvious
since in noninertial, that is, accelerated frames $c$ is no longer a
constant. In this sense, the vacuum admittance has a more fundamental
significance than the speed of light. The vacuum admittance can be
measured by a Weber-Kohlrausch type of experiment, e.g., see Raith
\cite{Raith} and Brown \cite{Brown}, the speed of light by the
well-known methods of Foucault or Fizeau,\footnote{For a history of
  the determination of the velocity of light before 1956, see, e.g.,
  Bergstrand \cite{Bergstrand}.} respectively (even though, strictly
speaking, the speed of light is put to a certain constant value in SI
since 1983, see \cite{NIST}).

Sommerfeld's fine structure constant, the dimensionless coupling
constant of the electromagnetic interaction, can be written as
\cite{Flowers,Zlatibor}
\begin{equation}\label{fine}
  \a_{\rm f}=\frac{Z_0}{2 R_{\rm K}}\,,
\end{equation}
where $R_{\rm K}$ is the quantum Hall resistance (von Klitzing
constant) associated with the quantum Hall effect. It has been shown
\cite{Rosenow} that $R_{\rm K}$, like the vacuum resistance
$Z_0$, is {\it not} influenced by the gravitational field. This
underlines the fundamental importance of $Z_0$ as well as that of
$R_{\rm K}$.

Let us now turn to {\it media}, namely to dielectric and magnetic
media that can be described by a {\it local and linear} constitutive
law.  Suppose we consider a spatially isotropic medium. In a frame
where the medium is at rest, we find
$\mathbf{D}=\varepsilon\ve_0\mathbf{E}$ and
$\mathbf{H}=\mathbf{B}/(\mu\mu_0)$, with the permittivity $\ve$ and
the permeability $\mu$ of the medium; both, $\ve$ and $\mu$, are
dimensionless and depend in general on the frequency of the wave
studied (``dispersion''). By the analogous arguments as above,
$\sqrt{\ve\ve_0/(\mu\mu_0)}$ is a 4-scalar in arbitrary coordinates
and $1/\sqrt{\ve\mu\ve_0\mu_0}$ a speed in inertial coordinates. The
absolute refractive index $n$ of a medium, see \cite{Born}, derives
from the latter expression as relation of the vacuum speed of light to
the speed in the medium. Accordingly, $n=\sqrt{\ve\mu}$.

Usually media, in particular crystalline media, behave
anisotropically. We could generalize the laws above by introducing
anisotropic permittivity and permeability tensors $\ve_{ab}$ and
$\mu_{ab}$, respectively, with $a,b,...=1,2,3$. But it is better to
start right away with a general {\it local and linear} constitutive
law in the context of a 4-dimensional representation of
electrodynamics (Sec.\ref{sec2}). This guarantees automatically
relativistic covariance, and the anisotropic laws mentioned will
emerge as special cases.

In this context, it turns out, see (\ref{constit1}), that the
corresponding 4-dimensional constitutive tensor
$\chi^{\lambda\nu\sigma\kappa}= -\chi^{\nu\lambda\sigma\kappa}=
-\chi^{\lambda\nu\kappa\sigma}$ has 36 independent components in
general, with $\lambda,\nu,...=0,1,2,3$. In the center of our present
paper is only one component of $\chi^{\lambda\nu\sigma\kappa}$, namely
its totally antisymmetric piece. It is the pseudoscalar
$\widetilde{\a}:= \widetilde{\epsilon}_{\lambda\nu\sigma\kappa}\,
\chi^{\lambda\nu\sigma\kappa}/4!$ (we sum over all repeated indices),
which can be formed from the constitutive tensor
$\chi^{\lambda\nu\sigma\kappa}$ with the help of the totally
antisymmetric Levi-Civita symbol
$\widetilde{\epsilon}_{\lambda\nu\sigma\kappa}$. We remind ourselves
that $ \widetilde{\epsilon}_{\lambda\nu\sigma\kappa}=+ 1$ or $=-1$
depending whether $\lambda\nu\sigma\kappa$ denotes an even or an odd
permutation of the numbers $0123$, respectively; it is zero otherwise,
see \cite{Sokol}. Then,
\begin{equation}\label{pss}
  \widetilde{\a}=\frac{1}{24}\left(\chi^{0123}- \chi^{0213} +
    \chi^{0312} -  \chi^{0132}    +-\dots\right)\,.
\end{equation}
Observing the antisymmetries of $\chi^{\lambda\nu\sigma\kappa}$,
we find
\begin{eqnarray}\label{pss1}
  \widetilde{\a}&=&\frac{1}{6}\left(\chi^{0123}+ \chi^{0231} +
    \chi^{0312}\nonumber\right.\\&&\left.\hspace{6pt}+
    \chi^{2301}+ \chi^{3102} +\chi^{1203} \right)\,.
\end{eqnarray}
The components $\chi^{0123}$...\ in the first line are related to the
magnetoelectric (ME) effect in an external {\it magnetic} field,
ME$_{\rm B}$, those in the second line to the
magnetoelectric effect in an external {\it electric} field, ME$_{\rm
  E}$. That $\widetilde{\a}$ is a pseudoscalar, indeed, can be seen
from the transformation properties of the quantities involved, see
\cite{Birkbook}. The dimension of $\widetilde{\a}$ is 1/{\it
  resistance}.

The constitutive tensor $\chi^{\lambda\nu\sigma\kappa}$ was seemingly
first introduced by Tamm \cite{Tamm}, see also Post \cite{Post}, and
later discussed by O'Dell \cite{O'Dell} in the context of
magnetoelectric media --- that is, media in which an {\it electric}
field $\mathbf{E}$ induces a {\it magnetic} excitation $\mathbf{H}$
and a magnetic field $\mathbf{B}$ and electric excitation
$\mathbf{D}$, see Fiebig \cite{Fiebig2} for a recent review. All four
cited authors assumed a further symmetry, namely
$\chi^{\lambda\nu\sigma\kappa}= \chi^{\sigma\kappa\lambda\nu}$
(``vanishing of the skewon part'', see \cite{Birkbook}). This symmetry
emerges, as soon as one stipulates that the constitutive relation can
be derived from a Lagrangian (thereby excluding irreversible
processes). Then, in particular, i.e., for
$\chi^{\lambda\nu\sigma\kappa}= \chi^{\sigma\kappa\lambda\nu}$,
\begin{equation}\label{pss2}
\widetilde{\a}=\frac 13\left(\chi^{0123}+
  \chi^{0231} + \chi^{0312}\right).
\end{equation}
Thus we need only three components of the constitutive tensor for the
determination of $\widetilde{\a}$ in the case of the vanishing skewon
part. Post argued \cite{Post} (not very convincingly, we should say)
that the pseudoscalar (\ref{pss2}) ought to vanish:
$\widetilde{\a}=0$.  This condition was dubbed ``Post constraint'' by
Lakhtakia \cite{Akhlesh1}. It was {\it not} assumed by O'Dell so that
he was left with 20+1 independent components of the constitutive
tensor, see \cite{O'Dell}, p.44. 

Later a fierce dispute arose about the Post constraint. The situation
was reviewed by Lakhtakia \cite{Akhlesh1,Akhlesh2}, de Lange and Raab
\cite{deLange}, Raab and Sihvola \cite{RaabSihvola1997}, Raab and de
Lange \cite{RaabBook}, Sihvola \cite{Ari1995}, Sihvola and Tretyakov
\cite{SihTre}, and in \cite{Postconstraint}, see also
\cite{measuringAxion,Optik}, where more references to the relevant
literature can be found. The evidence was mounting that there is no
reason to assume the validity of the Post constraint in general. This
point of view will be shown to be correct in this paper.

According to a theory of Dzyaloshinskii \cite{Dz1}, which was based on
an analysis of neutron scattering data, susceptibility measurements,
and symmetry (as derived in \cite{Brock}), a crystal of Chromium oxide
Cr$_2$O$_3$ is the substance par excellence for discovering the
magnetoelectric effect. In our paper, we consider only single domain
crystals.  In Sec.\ref{sec3}, we will describe Dzyaloshinskii theory
of Cr$_2$O$_3$ and we will determine the 4-dimensional pseudoscalar
$\widetilde{\a}$ of Cr$_2$O$_3$ in this framework. In Sec.\ref{sec4},
after a short introduction on dimensions and units, an overview will
be given over magnetoelectric experiments with Cr$_2$O$_3$.
Corresponding unpublished measurements by one of us (J.-P.R.) will be
presented in some detail.  In Sec.\ref{sec5}, we study the
experimentally determined magnetoelectric moduli of Cr$_2$O$_3$. Then,
we extract from the data, {\it for the first time,} the relativistic
pseudoscalar $\widetilde{\a}$ for Cr$_2$O$_3$.  It turns out to be
temperature dependent and is of the order of $\widetilde{\a}\approx
10^{-4}/Z_0$. This is a typical magnitude for magnetoelectric moduli
(Borovik-Romanov and Grimmer \cite{A}). Thus, it is small but
definitely {\it nonvanishing}.  This proves experimentally that the
Post constraint is ruled out as a generally valid law. In
Sec.\ref{sec6}, we show that the pseudoscalar (or axion) piece of the
magnetoelectric susceptibility of Cr$_2$O$_3$ violates parity and time
inversion invariance and doesn't contribute to the electromagnetic
energy. In Sec.\ref{sec7}, we mention other substances that, besides
Cr$_2$O$_3$, carry an axion piece, and in Sec.\ref{sec8}, we will
discuss the implications of our result to other disciplines within
physics and electrical engineering.

In this article, we will base our considerations on the 4-dimensional
tensor analytical formalism as described by
Post\footnote{Incidentally, Post's book, as a Dover edition, is easily
  available.} \cite{Post}, which can be understood by experimentalists
and theoreticians alike. Accordingly, the time coordinate $x^0=t$ has
the dimension of `time' whereas the spatial coordinates $x^1,x^2,x^3$
are related to the dimension of 'length'.  Only intermittently we will
mention the formalism of exterior differential forms, as it is used in
the book \cite{Birkbook}. The equations in our paper are quantity
equations that are valid for an arbitrary system of units.  However,
if in Secs.\ref{sec4} and \ref{sec5} we turn to experiments, then we
use SI and sometimes Gaussian units, which are widespread in the
literature.

\section{Local and linear magnetoelectric media, the pseudo\-scalar
  $\tilde{\a}$}\label{sec2}

\subsection{Excitation and field strength}

Post \cite{Post} represents the 4-dimensional electromagnetic field
tensors according to
\begin{eqnarray}\label{frakG}
  \mathfrak{ G}^{\mu\nu}=-\mathfrak{ G}^{\nu\mu}&=& \left(\begin{array}{cccc}
      0\hspace{5pt} & D_1 & D_2 & D_3\\ -D_1 &0\hspace{5pt} & H_3 &
      -H_2\\ -D_2 & -H_3 &0\hspace{5pt} & H_1\\ -D_3 & H_2 & -H_1
      &0\hspace{5pt}
  \end{array}\right)\,,\\&&\cr
\label{F}
F_{\mu\nu}=-F_{\nu\mu}&=&\left(\begin{array}{cccc}0 &
    -E_1 & -E_2 & -E_3\\ E_1 &0\hspace{5pt} & B_3 & -B_2\\ E_2 & -B_3
    &0\hspace{5pt} & B_1\\ E_3 & B_2 & -B_1 &0\hspace{5pt}
  \end{array}\right).
\end{eqnarray}
The coordinate indices $\mu,\nu,...$ run from 0 to 3. We have to
distinguish carefully between the upper and lower indices for reasons
of general 4-dimensional invariance. We use Post's conventions
throughout, unless indicated otherwise.

The field $\mathfrak{G}^{\mu\nu}$, the electromagnetic {\it excitation},
represents a tensor density of weight $+1$. $F_{\mu\nu}$ is a tensor.
In order to transform $\mathfrak{G}^{\mu\nu}$ to a pseudo tensor with
lower indices, we introduce (Einstein's summation convention assumed)
\begin{equation}\label{Gdual}
  \widetilde{G}_{\mu\nu}:=\frac
  12\,\widetilde{\epsilon}_{\mu\nu\kappa\lambda}
  \mathfrak{G}^{\kappa\lambda}=- \widetilde{G}_{\nu\mu}\,,
\end{equation}
with the totally antisymmetric Levi-Civita symbol
$\widetilde{\epsilon}_{\mu\nu\kappa\lambda}=0,\pm 1$. Here
$\widetilde{\epsilon}_{\mu\nu\kappa\lambda}$ is a pseudo tensor
density of weight $-1$. We use
$\widetilde{\epsilon}_{\mu\nu\kappa\lambda}$ for the dualization in
(\ref{Gdual}), since this relation is invariant under proper as well
as under improper transformations (i.e., those that include coordinate
reflections). We will denote pseudo tensors by a tilde throughout our
article.
By simple algebra we can construct the corresponding matrix for
$\widetilde{G}_{\mu\nu}$:
\begin{equation}\label{Gtwiddle}
  \widetilde{G}_{\mu\nu}=-\widetilde{G}_{\nu\mu}
  =\left(\begin{array}{cccc}
      0\hspace{5pt} & H_1 & H_2 & H_3\\ -H_1 &0\hspace{5pt} & D_3 &
      -D_2\\ -H_2 & -D_3 &0\hspace{5pt} & D_1\\ -H_3 & D_2 & -D_1
      &0\hspace{5pt}
  \end{array}\right).
\end{equation}
Whereas $F_{\mu\nu}$, the electromagnetic {\it field strength}, is a
tensor, $\widetilde{G}_{\mu\nu}$ is a {\it pseudo} tensor, i.e., they
behave differently under a reflection $x^0\rightarrow -x^0$ or
$x^1\rightarrow -x^1$ etc.  In the language of differential forms, the
electromagnetic field can be represented by two two-forms, the
excitation $ \widetilde{G}=\widetilde{G}_{\mu\nu}\,dx^\mu\wedge
dx^\nu/2$ and the field strength $F=F_{\mu\nu}\,dx^\mu\wedge
dx^\nu/2$, respectively. The $\widetilde{G}$ is a pseudo 2-form, also
called {\it twisted} (or odd) 2-form, the $F$ is an untwisted (or
even) 2-form.

\subsection{The Maxwell equations}

The Maxwell equations in premetric form are \cite{Post}:
\begin{equation}\label{Maxwell1}
  \partial_\nu\mathfrak{G}^{\mu\nu}=\mathfrak{J}^\mu\,,\qquad \partial_\mu
  F_{\nu\lambda}+\partial_\nu F_{\lambda\mu}+ \partial_\lambda
  F_{\mu\nu}=0\,.
\end{equation}
If we introduce the dual of the electric current density,
$\widetilde{J}_{\mu\nu\lambda}:=\widetilde{\epsilon}_{\mu\nu\lambda\kappa}
\,\mathfrak{J}^\kappa$, then the inhomogeneous equation can be
transformed into the equivalent equation
\begin{equation}\label{Maxwell2}
  \partial_\mu\widetilde{G}_{\nu\lambda}+\partial_\nu
  \widetilde{G}_{\lambda\mu}+ \partial_\lambda
  \widetilde{G}_{\mu\nu}=\widetilde{J}_{\mu\nu\lambda}\,.
\end{equation}
In a more condensed way, we may also write the Maxwell equations as
\begin{equation}\label{Maxwell3}
  \partial_{[\mu}\widetilde{G}_{\nu\lambda]}=\frac 13
  \widetilde{J}_{\mu\nu\lambda}\,,\qquad
  \partial_{[\mu}F_{\nu\lambda]}=0\,,
\end{equation}
where $[\mu\nu\lambda]=\frac 16(\mu\nu\lambda
-\nu\mu\lambda+\lambda\mu\nu\mp \cdots)$, i.e., a plus (minus) sign
occurs before even (odd) permutations.  In the differential form
language, the Maxwell equations (\ref{Maxwell3}) read
\begin{equation}\label{Max-df}
  d\widetilde{G}=\widetilde{J}\,, \qquad dF=0\,,
\end{equation}
with the 3-form of the electric current $\widetilde{J}=\frac {1}{6}
\widetilde{J}_{\mu\nu\lambda}\,dx^\mu\wedge dx^\nu\wedge dx^\lambda$.

\subsection{Constitutive relation}

The system of the $4+4$ Maxwell equations (\ref{Maxwell1}) for the
$6+6$ independent components of the electromagnetic field
$\mathfrak{G}^{\mu\nu}$ and $F_{\lambda\rho}$ is evidently
underdetermined. To complete this system, a constitutive relation of
the form
 \begin{equation}\label{Pconst0}
   \mathfrak{G}^{\mu\nu}= \mathfrak{G}^{\mu\nu}(F_{\lambda\rho})
 \end{equation}
 has to be assumed. The constitutive relation (\ref{Pconst0}) is
 independent of the Maxwell equations and its form can be determined
 by using experimental results. For {\it vacuum,} the constitutive
 relation (``spacetime relation'') is simple,
 \begin{equation}\label{vacuum2} \mathfrak{G}^{\lambda\nu}=
   Y_0\sqrt{-g}F^{\lambda\nu}\,,
\end{equation}
with $g:=\det g_{\rho\sigma}\ne 0$ and
$F^{\mu\nu}:=g^{\mu\a}g^{\nu\b}\,F_{\a\b}$. Here $g_{\mu\nu}$ are the
covariant components of the metric of spacetime with signature
$(-++\,+)$, its contravariant components $g^{\lambda\nu}$ can be
determined via $g_{\mu\lambda}\,g^{\lambda\nu}= \delta_{\mu}^\nu$. For
the excitation pseudo tensor, we have $\widetilde{G}_{\mu\nu}=Y_0\,
\widetilde{\epsilon}_{\mu\nu\sigma\tau}\sqrt{-g}\,F^{\sigma\tau}/2$
and, in exterior calculus, with the Hodge $^\star$ operator,
$\widetilde{G}=Y_0\,^\star\! F$.

Turning to general {\it magnetoelectric} media, we assume with Tamm
\cite{Tamm}, see also Post \cite{Post}, the most general
local\footnote{Kamenetskii \cite{Kamenetskii} gave a general
  discussion of linear (`bianisotropic') media that also takes
  {\it non-local} effects into account. Ascher \cite{Ascher74}
  investigated relativistic symmetries in the context of the
  magnetoelectric effect.} and linear
homogeneous relation
\begin{equation}\label{constit1}
  \mathfrak{G}^{\lambda\nu}=\frac
  12\,\chi^{\lambda\nu\sigma\kappa}F_{\sigma\kappa}\,,
\end{equation}
where $\chi^{\lambda\nu\sigma\kappa}$ is a {\it constitutive tensor
  density} of rank 4 and weight $+1$, with the dimension
$[\chi]=1/resistance$. Since both $ \mathfrak{G}^{\lambda\nu}$ and
$F_{\sigma\kappa}$ are antisymmetric in their indices, we have
$\chi^{\lambda\nu\sigma\kappa}=-\chi^{\lambda\nu\kappa\sigma}=
-\chi^{\nu\lambda\sigma\kappa}$.  An antisymmetric pair of indices
corresponds, in four dimensions, to six independent components. Thus,
the constitutive tensor can be considered as a $6\times 6$ matrix with
36 independent components.

A $6\times 6$ matrix can be decomposed in its tracefree symmetric part
(20 independent components), its antisymmetric part (15 components),
and its trace (1 component). On the level of
$\chi^{\lambda\nu\sigma\kappa}$, this {\it decomposition} is reflected
in
\begin{eqnarray}\label{chidec}
  \chi^{\lambda\nu\sigma\kappa}&=&\,^{(1)}\chi^{\lambda\nu\sigma\kappa}+
  \,^{(2)}\chi^{\lambda\nu\sigma\kappa}+
  \,^{(3)}\chi^{\lambda\nu\sigma\kappa}\,.\\ \nonumber 36
  &=&\hspace{15pt} 20\hspace{15pt}\oplus \hspace{15pt}15\hspace{15pt}
  \oplus \hspace{25pt}1\,.
\end{eqnarray}
The third part, the {\it axion} part, is totally antisymmetric and as
such proportional to the Levi-Civita symbol, $
^{(3)}\chi^{\lambda\nu\sigma\kappa}:= \chi^{[\lambda\nu\sigma\kappa]}
=\widetilde{\a}\, \widetilde{\epsilon}^{\lambda\nu\sigma\kappa}$. The
second part, the {\it skewon} part, is defined according to $
^{(2)}\chi^{\mu\nu\lambda\rho}:=\frac 12(\chi^{\mu\nu\lambda\rho}-
\chi^{\lambda\rho\mu\nu})$.  If the constitutive equation can be
derived from a Lagrangian, which is the case as long as only
reversible processes are considered, then
$^{(2)}\chi^{\lambda\nu\sigma\kappa}=0$. The {\it principal} part
$^{(1)}\chi^{\lambda\nu\sigma\kappa}$ fulfills the symmetries $
^{(1)}\chi^{\lambda\nu\sigma\kappa}=
{}^{(1)}\chi^{\sigma\kappa\lambda\nu}$ and
$^{(1)}\chi^{[\lambda\nu\sigma\kappa]}=0$.  The constitutive relation
now reads
\begin{equation}\label{constit7}
  { \mathfrak{G}^{\lambda\nu}=\frac
    12\left({}^{(1)}{\chi}^{\lambda\nu\sigma\kappa}+
      {}^{(2)}{\chi}^{\lambda\nu\sigma\kappa} +\widetilde{\a}\,
      \widetilde{\epsilon}^{\lambda\nu\sigma\kappa}\right)F_{\sigma\kappa}\,.}
\end{equation}

In order to compare this with experiments, we have to split
(\ref{constit7}) into time and space parts. As shown in
\cite{Postconstraint} in detail, we can parametrize the {\it
  principal} part by the 6 permittivities $\varepsilon^{ab}=
\varepsilon^{ba}$, the 6 permeabilities $\mu_{ab}=\mu_{ba}$, and the 8
magnetoelectric pieces $\g^a{}_b$ (its trace vanishes, $\g^c{}_c=0$)
and the {\it skewon} part by the 3 permittivities $n_a$, the 3
permeabilities $m^a$, and the 9 magnetoelectric pieces $s_a{}^b$. Then,
the constitutive relation (\ref{constit7}) can be rewritten as
\begin{eqnarray}\label{explicit'}
  {D}^a\!&=\!&\left( \varepsilon^{ab}\hspace{4pt} - \,
    \epsilon^{abc}\,n_c \right)E_b\,+\left(\hspace{9pt} \gamma^a{}_b +
    s_b{}^a - \delta_b^a s_c{}^c\right) {B}^b +
  \widetilde{\alpha}\,B^a \,, \\ {H}_a\!&=\!&\left( \mu_{ab}^{-1}
    - \hat{\epsilon}_{abc}m^c \right) {B}^b +\left(- \gamma^b{}_a +
    s_a{}^b - \delta_a^b s_c{}^c\right)E_b -
  \widetilde{\alpha}\,E_a\,.\label{explicit''}
\end{eqnarray}
Here $\epsilon^{abc}= \hat{\epsilon}_{abc}=\pm 1,0$ are the
3-dimensional Levi-Civita symbols. As can be seen from our derivation,
$\widetilde{\alpha}$ is a 4-dimensional pseudo (or axial) scalar,
whereas $s_c{}^c$ is only a 3-dimensional scalar. The cross-term
$\gamma^a{}_b$ is related to the Fresnel-Fizeau effects.  The skewon
contributions $m^c,n_c$ are responsible for electric and magnetic
Faraday effects, respectively, whereas the skewon terms $s_a{}^b$
describe optical activity. Equivalent constitutive relations were
formulated by Serdyukov et al.\ \cite{Serdyukov}, p.86, and studied in
quite some detail.

According to Post \cite{Post}, the pseudoscalar $\widetilde{\a}$
should vanish for the vacuum and for all media. We will show in the
next section that, in general, this is not the case. For Cr$_2$O$_3$
the pseudoscalar $\widetilde{\a}$ turns out to be finite.

\section{The antiferromagnet Cr$_2$O$_3$ and the theory of\\ 
Dzyaloshinskii}\label{sec3}

On the basis of neutron scattering data \cite{Brock} and
susceptibility measurements \cite{McGuire} of the antiferromagnetic
chromium sesquioxide\footnote{A corresponding mineral with about 94\%
  Cr$_2$O$_3$ is called Escolaite.} Cr$_2$O$_3$, Dzyaloshinskii
\cite{Dz1} was able to establish the magnetic symmetry class
$\overline{3}{}'m'$ of the Cr$_2$O$_3$ crystals.  In accordance with
these results, Dzyaloshinskii developed, by starting from a
thermodynamic potential quadratic and bilinear in $\mathbf{E}$ and
$\mathbf{H}$, as foreseen by Landau \& Lifshitz \cite{LL}, a theory
for the electromagnetic constitutive relations for Cr$_2$O$_3$.  We
write them here as {\it quantity equations} that are valid in an
arbitrary system of units:\footnote{For the microscopic origin of such
  relations, compare Gehring \cite{Gehring} and Tol\'edano
  \cite{Toledano}.}
\begin{eqnarray}\label{DH1}
  D_x &=& \varepsilon_{\bot} \varepsilon_0 E_x
  +\alpha_{\bot}\sqrt{\varepsilon_0\mu_0} H_x\,,\\ \label{DH2} D_y &=&
  \varepsilon_{\bot} \varepsilon_0 E_y +\alpha_{\bot}
  {\sqrt{\varepsilon_0\mu_0}} H_y\,,\\ \label{DH3} D_z &=&
  \varepsilon_{||}\,\varepsilon_0 E_z +\alpha_{||}\,{\sqrt{
      \varepsilon_0\mu_0}} H_z\,,
\end{eqnarray}
and
\begin{eqnarray}\label{BE1}
  B_x &=& \mu_{\bot}\mu_0 H_x +\alpha_{\bot}\sqrt{\varepsilon_0\mu_0}
  E_x\,,\\\label{BE2} B_y &=& \mu_{\bot}\mu_0 H_y
  +\alpha_{\bot}\sqrt{\varepsilon_0\mu_0} E_y\,,\\\label{BE3} B_z &=&
  \mu_{||}\,\mu_0 H_z +\alpha_{||}\,\sqrt{\varepsilon_0\mu_0} E_z\,.
\end{eqnarray}
The $z$-axis is parallel to the optical axis of Cr$_2$O$_3$. Remember
that $\sqrt{\varepsilon_0\mu_0}=1/c$ and $\sqrt{\ve_0/\mu_0} =1/Z_0$,
with $c$ as velocity of light and $Z_0$ as vacuum impedance. Here we
have permittivities parallel and perpendicular to the z-axis of the
crystal, namely $\varepsilon_{||},\varepsilon_{\bot}$, analogous
permeabilities $ \mu_{||}, \mu_{\bot}$ and magnetoelectric moduli
$\alpha_{||},\alpha_{\bot}$. Note that all these moduli are
dimensionless, and this is true for all systems of units.  Our
dimensionless $\alpha$'s are different from the ones used by
experimentalists and theoreticians up to now. We will discuss the
transition to different systems of units in Sec.\ref{sec41} below.

As we can see from (\ref{frakG}) and (\ref{constit1}), we have to get
$(\mathbf{D},\mathbf{H})$ on the left hand side and
$(\mathbf{E},\mathbf{B})$ on the right hand side in order to end up
with a constitutive law that is written in a relativistically
covariant form. For this purpose, we resolve the last three equations
with respect to $H$:
\begin{eqnarray}\label{HB1} 
  H_x &=& \frac{1}{\mu_\bot{\mu_0}} B_x
  -\frac{\alpha_{\bot}}{\mu_\bot}\sq2 E_x\,,\\ \label{HB2} H_y &=&
  \frac{1}{\mu_\bot{\mu_0}} B_y -\frac{\alpha_{\bot}}{\mu_\bot}\sq2
  E_y\,,\\ 
  \label{HB3} H_z &=& \frac{1}{\mu_{||}{\mu_0}} B_z
  -\frac{\alpha_{||}}{\mu_{||}}\,\sq2 E_z\,.
\end{eqnarray}
On substitution into (\ref{DH1}) to (\ref{DH3}), we find,
\begin{eqnarray}\label{Dcon1}
  D_x &=& \left(\varepsilon_{\bot}-\frac{\a_\bot^2}{\mu_\bot}\right)
  \varepsilon_0 E_x +\frac{\alpha_{\bot}}{\mu_\bot}\sq2 B_x\,,\\ D_y
  &=&\left( \varepsilon_{\bot}-\frac{\a_\bot^2}
    {\mu_\bot}\right)\varepsilon_0 E_y +\frac{\alpha_{\bot}}
  {\mu_\bot}\sq2 B_y\,,\\ D_z &=&\left(
    \varepsilon_{||}-\frac{\a_{||}^2}{\mu_{||}}\right)\,
  \varepsilon_0E_z +\frac{\alpha_{||}}{\mu_{||}}\,\sq2
  B_z\,.\label{Dcon3}
\end{eqnarray}

Now we have to compare with the local and linear constitutive relation
(\ref{explicit'}) and (\ref{explicit''}). Since Dzyaloshinskii assumed
that his constitutive relations can be derived from a Hamiltonian, it
is clear that the {\it skewon piece} with its 15 independent
components {\it vanishes} identically, see \cite{Birkbook},
Eq.(D.1.44). But this can be also read off from comparing
(\ref{explicit'}) and (\ref{explicit''}) with (\ref{HB1}) to
(\ref{Dcon3}). The skewon pieces $n_c$ and $m^c$ must be zero, since the
$D_a$ in (\ref{Dcon1}) to (\ref{Dcon3}) are proportional to $E_a$, and
the $H_a$ in (\ref{HB1}) to (\ref{HB3}) are proportional to $B_a$. A
similar consideration shows that $s_a{}^b=0$, since $s_a{}^b$ only
provides off-diagonal pieces. Consequently, eqs.(\ref{explicit'}) and
(\ref{explicit''}) reduce to
\begin{eqnarray}\label{explicit3}
  {D}^a\!&=\!& {\varepsilon^{{ab}}}\,E_b + {\gamma^a{}_b}\, {B}^b +
  {\widetilde{\a}}\,B^a \,,\\ {H}_a\!  &=\!  & { \mu_{ab}^{-1}} {B}^b - {
    \gamma^b{}_a}E_b - {\widetilde{\a}}\,E_a\,,\label{explicit4}
\end{eqnarray}
with 21 independent moduli. The permittivity matrix $\varepsilon^{ab}$
and the impermeability matrix $\mu_{ab}^{-1}$ are both symmetric and
possess 6 independent components each, the magnetoelectric cross-term
with $\gamma^a{}_b$, which is tracefree, $\gamma^c{}_c=0$, has 8
independent components. The 4-dimensional pseudo scalar (we call it
also the axion parameter) represents 1 component.

By comparing (\ref{explicit3}) and (\ref{explicit4}) with the above
equations (\ref{Dcon1}) to (\ref{Dcon3}) and (\ref{HB1}) to
(\ref{HB3}), we can read off the permittivity
\begin{equation}\label{perm1}
  \varepsilon^{ab}= \varepsilon_0\begin{pmatrix}
    \varepsilon_\bot-\frac{\a_\bot^2}{\mu_\bot}&0&0\cr 0&
    \varepsilon_\bot-\frac{\a_\bot^2}{\mu_\bot}&0\cr 0&0&
    \varepsilon_{||}-\frac{\a_{||}^2}{\mu_{||}} \end{pmatrix}
\end{equation}
and the impermeability
\begin{equation}\label{perm2}
  \mu^{-1}_{ab}= \mu_0^{-1}\begin{pmatrix} \mu^{-1}_\bot&0&0\cr 0&
  \mu^{-1}_\bot&0\cr 0&0& \mu^{-1}_{||}\end{pmatrix}\,.
\end{equation}
For the magnetoelectric cross-terms, we have
\begin{eqnarray}\label{cross1}
  \hspace{9pt} \gamma^x{}_x \, {B}^x +
  \widetilde{\a}\,B^x & =&\frac{\alpha_{\bot}} {\mu_\bot}\sq2 B_x\,,\\ 
  \hspace{9pt} \gamma^y{}_y \, {B}^y +
  \widetilde{\a}\,B^y &=& \frac{\alpha_{\bot}}{\mu_\bot}\sq2 B_y\,,\\ 
  \hspace{9pt} \gamma^z{}_z \,{B}^z +
  \widetilde{\a}\,B^z &=&\frac{\alpha_{||}}{\mu_{||}}\,\sq2 B_z\,,
\label{cross1'}
\end{eqnarray}
and
\begin{eqnarray}\label{cross2}
  - \gamma^x{}_x\,E_x - \widetilde{\a}\,E_x
  &=& -\frac{\alpha_{\bot}}{\mu_\bot}\sq2 E_x\,,\\ - \gamma^y{}_y \,E_y 
- \widetilde{\a}\,E_y &=&
  -\frac{\alpha_{\bot}}{\mu_\bot}\sq2 E_y\,,\\ - \gamma^z{}_z \,E_z 
- \widetilde{\a}\,E_z &=&
  -\frac{\alpha_{||}}{\mu_{||}}\,\sq2 E_z\,.\label{cross2'}
\end{eqnarray}
Note that in the Cartesian coordinates used by Dzyaloshinskii we have
$B^x=B_x$ etc., since the spatial metric is Euclidean with signature
$(+++)$. Thus, we are left with
\begin{eqnarray}\label{cross3a}
  \gamma^x{}_x + \widetilde{\a} & =&\frac{\alpha_{\bot}} {\mu_\bot}\sq2 \,,\\ 
  \label{cross3b} \gamma^y{}_y + \widetilde{\a} &=&
  \frac{\alpha_{\bot}}{\mu_\bot}\sq2 \,,\\ \label{cross3c} \gamma^z{}_z +
  \widetilde{\a} &=&\frac{\alpha_{||}}{\mu_{||}}\sq2 \,.
\end{eqnarray}
One of the triplets of equations (\ref{cross1}) to (\ref{cross1'}) and
(\ref{cross2}) to (\ref{cross2'}) is redundant because of the
vanishing of the skewon piece of $\chi^{\lambda\nu\sigma\kappa}$

The matrix $\gamma$ is traceless: $ \gamma^x{}_x + \gamma^y{}_y +
\gamma^z{}_z=0$. If we add up all three equations, we find for the
pseudoscalar (or axion) piece
\begin{equation}\label{axion}
  \boxed{  \widetilde{\a}= \frac 13\left(2\,\frac{\a_\bot}{\mu_\bot}+ 
      \frac{\a_{||}}{\mu_{||}} \right)\sq2 \,.}
\end{equation}
Resubstituted into (\ref{cross3a}), (\ref{cross3b}), and
(\ref{cross3c}), the magnetoelectric $\gamma$ matrix becomes
\begin{equation}\label{gamma}
  \gamma^a{}_b= \frac 13\left(\frac{\a_\bot}{\mu_\bot}- 
\frac{\a_{||}}{\mu_{||}} \right)\sq2 
  \begin{pmatrix}1&0&0\cr
    0&1&0\cr
    0&0&\hspace{-6pt}-2 \end{pmatrix}\,,
\end{equation}
that is, it has only nonvanishing diagonal components! The
magnetoelectric matrix $\g^a{}_b$ as well as the pseudo scalar
$\widetilde{\a}$ carry the dimension of 1/resistance.  Conventionally,
in the ``magnetoelectric literature'' the $\g$-matrix and
$\widetilde{\a}$ are collected in the ``relativistic'' matrix
\begin{equation}\label{alpha}
^{\rm rel}\a^a{}_b:=\g^a{}_b+\widetilde{\a}\,\delta^a_b=\sq2\begin{pmatrix}
\frac{\a_\bot}{\mu_\bot}&0&0\\
0&\frac{\a_\bot}{\mu_\bot}&0\\
0&0&\frac{\a_{||}}{\mu_{||}}\end{pmatrix}\,.
\end{equation}
It is called relativistic, since it occurs in the context of the
relativistic $(\mathbf{E},\mathbf{B})$ system, see (\ref{explicit3}).

Since there are doubts in the literature about the correctness of
Dzyaloshinskii's theory, see Lakhtakia \cite{Akhlesh1}, it is
important to note that O'Dell \cite{O'Dell}, pp.\ 115/116, and Janner
\cite{Janner}, p.205, (see also Rado \& Folen \cite{RadoFolen62} and
O'Dell \cite{O'Dell1966}) analyzed the crystal structure of
Cr$_2$O$_3$ and determined the form the matrices $\ve^{ab}$,
$\mu_{ab}^{-1}$, and $^{\rm rel}\a^a{}_b$ ought to possess. They found
\begin{equation}\label{structure}
\ve^{ab}\sim
\mu_{ab}^{-1}\sim\,^{\rm rel}\a^a{}_b\sim
\begin{pmatrix}
\bullet & . & .\\
  . & \bullet & .\\
 . & . & *\end{pmatrix}\,,
\end{equation}
where nonvanishing entries are denoted by $\bullet$ and *,
respectively. A comparison with (\ref{perm1}), (\ref{perm2}), and
(\ref{gamma}) confirms Dzyaloshinskii's theory.

Summing up: the nonvanishing magnetoelectric moduli for Cr$_2$O$_3$
can be determined with the help of eqs.\ (\ref{axion}) and
(\ref{gamma}). Let us stress that $\ve_{||},\ve_\bot,\mu_{||},$
$\mu_\bot, \a_{||}$, and $\a_\bot$, according to their definitions
(\ref{DH1}) to (\ref{BE3}), are measured in an external $\mathbf{E}$
and/or an external $\mathbf{H}$ field.

\section{ Magnetoelectric experiments 
with Cr$_2$O$_3$}\label{sec4}

\subsection{Dimensions and units}\label{sec41}

Since in the literature, which we need for extracting data, different
systems of units are used, we want to underline again that usually all
our equations are {\it quantity equations,} which are valid in all
systems of units; only in the ``experimental'' Secs.\ref{sec4} and
\ref{sec5}, we will turn to specific systems of units and some
equations may be unit-dependent, see also Rivera \cite{Rivera3} in
this context. We will go into some detail here, since these questions
often lead to misunderstandings between theoreticians and
experimentalists.

A {\it physical quantity} $Q$ is given by
\begin{equation}\label{1}
  Q=\{Q\}\,[Q]\,.
\end{equation}
Here $\{Q\}$ is a numerical value and $[Q]$ the physical dimension of
the quantity $Q$. For instance, we have for a (1-dimensional) {\it
  displacement} $s$,
\begin{equation}\label{2}s=\{s\}\,[s]=\{s\}\,\mbox{length}\,,
\end{equation}
where length is the dimension of $s$. So far, all of the equations in
Secs.\ref{sec1} to \ref{sec3} are quantity equations. They interrelate
``physical quantities'' that consist of numbers and dimensions. Like
in (\ref{1}) and (\ref{2}), they are totally independent of any system
of units. In many papers and books, the equations are only valid in
one system of units, they are {\it numerical} equations, like in
Jackson's book on electrodynamics \cite{Jackson}, for
example.\footnote{Hence on each 2nd page in the top line Jackson marks
  whether he is in SI or in the Gaussian system.} In our paper, as in
Post \cite{Post} or in \cite{Birkbook}, the equations are quantity
equations and are {\it always} valid, independent of the units chosen.

In a second step, if we relate our equations to measured values, we
need a {\it system of units}. Then, for the example above,
\begin{eqnarray}\label{5}
  s&=&15\,m =15 00\,cm\approx 45\, f\!eet=...\nonumber\\
  &=&\{s\}'\,m=\{s\}'''\,cm=\{s\}''\,f\!eet=...
\end{eqnarray}
The numerical value $\{s\}$ depends on the unit chosen. In fact,
\begin{eqnarray}\label{6}
  \frac{\{s\}'}{\{s\}''}=\frac{[s]''}{[s]'}\,,\quad 
  \frac{\{s\}'}{\{s\}'''}=\frac{[s]'''}{[s]'}\,,
\end{eqnarray}
etc., that is, we have {\it reciprocal proportionality}. This is one
of the fundamental laws of dimensional theory. The physical quantity
$s$ is {\it invariant}, i.e., it doesn't change, but its numerical
value $\{s\}$ may change according to the choice of the unit.

In the center of our interest is the pseudoscalar $\widetilde{\a}$.
According to (\ref{explicit4}), it has the dimension
\begin{equation}\label{dim1}
  [\widetilde{\a}]=\frac{[H_a]}{[E_a]}=
  \frac{\mbox{current}}{\mbox{length}}\times
  \frac{\mbox{length}}{\mbox{voltage}}=\frac{1}{\mbox{resistance}}\,.
\end{equation}
Moreover, the permittivity matrix $\ve^{ab}$, according to
(\ref{explicit3}), has the dimension
\begin{equation}\label{dim2}
  [\ve^{ab}]=\frac{[D^a]}{[E_a]}=\frac{\mbox{charge}}{\mbox{area}}\times
  \frac{\mbox{length}}{\mbox{voltage}}=\frac{\mbox{1}}{\mbox{velocity}}
  \times\frac{1}{\mbox{resistance}}\,
\end{equation}
and the impermeability matrix $\mu_{ab}^{-1}$, according to
(\ref{explicit4}),
\begin{equation}\label{dim3}
  [\mu_{ab}^{-1}]=\frac{[H_a]}{[B^a]}=\frac{\mbox{current}}
{\mbox{length}}\times
  \frac{\mbox{area}}{\mbox{voltage}\times\mbox{
      time}}=\mbox{velocity}\times\frac{1}{\mbox{resistance}}\,.
\end{equation}
In particular, we have 
\begin{equation}\label{dim4}
[\ve_0]=\frac{\mbox{1}}{\mbox{vel.}}
\times\frac{1}{\mbox{resist.}}\,,\;
\,\quad [\mu_0^{-1}] =\mbox{vel.}\times\frac{1}{\mbox{resist.}}\,,\quad
\left[\sq2\right]=\frac{1}{\mbox{resist.}}\,.
\end{equation}
Accordingly, we can summarize these considerations in
\begin{equation}\label{dim5}
  \frac{[\ve^{ab}]}{[\ve_0]}=\frac{[\mu_{ab}^{-1}]}{[\mu_0^{-1}]}=
  \frac{[\g^a{}_b]}{[\sqrt{\ve_0/\mu_0}]}=\frac{[\widetilde{\a}]}
  {[\sqrt{\ve_0/\mu_0}]}=1\,.
\end{equation}
These quotients are dimensionless in {\it all} systems of units. Here
$\ve_0$ and $\mu_0$ are universal constants. They have in different
systems of units different {\it numerical values}. We have
$\sqrt{\ve_0\mu_0}=1/c$ and, in particular,
\begin{eqnarray}\label{dim6}
  \sq2&=&\frac{1}{Z_0}\approx \frac{1}{376.73\;\Omega}\quad\mbox{in SI,}\\
  \sq2&=&\frac{c}{4\pi}\quad\mbox{in Gaussian units,}\nonumber\\
  \sq2&=&c\quad\mbox{in Heaviside-Lorentz (``rationalized
    Gaussian'') units.}\nonumber
\end{eqnarray}
Therefore, if you see $\varepsilon_0$ and $\mu_0$ in our equations,
like in (\ref{DH1}) to (\ref{BE3}), it does {\em not} mean that we are
in SI, but rather that we use {quantity equations} with correct
dimensions.

The thermodynamic potential (an energy density) relevant for the
magnetoelectrical effect contains cross terms between electric and
magnetic fields. As a quantity equation, it reads, restricting
ourselves to the linear regime,
\begin{eqnarray}\label{potential0}
  - g(\mathbf{E},\mathbf{H}; T)& =&
  \cdots+\underbrace{\a^{ab}}_{\rm t/\ell}E_a
  H_b=\cdots+\sqrt{\ve_0\mu_0}
  \underbrace{\a_*^{ab}}_{\rm dim.-less}E_aH_b\,.
\end{eqnarray}
This linear part of a power series development is valid for ``small''
values of the $E_a$ and $H_b$ fields, at least relatively to the
internal crystal fields. The $\a_\bot$ and $\a_{||}$ of Sec.\ref{sec3}
are components of the matrix $\a_*^{ab}$. Hence, strictly speaking, we
should have put a star to all of them. However, for convenience we
dropped these stars.

We will concentrate here on SI and on the Gaussian system. In the
Gaussian system of units, a mixed system consisting of electrical
(electrostatic) and magnetic cgs-units, see, e.g., Sommerfeld
\cite{Sommerfeld}, Panofsky \& Phillips \cite{Panofsky}, or Jackson
\cite{Jackson}, we have the following field redefinitions:
\begin{eqnarray}\label{redef}
  ^{\rm G}\!E_a=E_a\,,\quad ^{\rm G}\!D_a=4\pi D_a\,,\quad^{\rm
    G}\!H_a=\frac{4\pi}{c}
  H_a\,,\quad ^{\rm G}\!B_a=cB_a\,.
\end{eqnarray}
The speed of light $c$ is instrumental for making the dimensions of $
^{\rm G}\!E_a$ and $ ^{\rm G}\!B_a$ equal to each other, $[ ^{\rm
  G}\!E_a]=[^{\rm G}\!B_a]$, and, similarly, $[ ^{\rm G}\!D_a]=[^{\rm
  G}\!H_a]$. The $4\pi$ removes this factor from the Coulomb law. With
the field redefinitions (\ref{redef}) and with the convention in the
Gaussian system $ {}^{\rm G}\!{{\a}}^{ab}:=c\,{{\a}}^{ab}$,
Eq.(\ref{potential0}) can be rewritten as (see Landau-Lifshitz
\cite{LL} and Dzyaloshinskii \cite{Dz1} for the Gaussian
system)\footnote{Sometimes strange units are taken for $\a$. Often, in
  papers concerning composite materials, $\frac{V}{cm}/O\!e$ is used,
  see, e.g.\ \cite{Fiebig2}.  Needless to say that our {\em
    dimensionless} $ ^{\rm SI}\!\a_*^{ab}$ is to be preferred (or the
  dimensionless $^{\rm G}\a^{ab}$ or $^{\rm SI}\!{\a}^{ab}$ in
  $s/m$).}
\begin{eqnarray}\label{potential0'}
  - g(\mathbf{E},\mathbf{H}; T)& =&
  \cdots +
  \underbrace{ {}^{\rm SI}\!{\a}^{ab}}_{\rm t/\ell}{}^{\rm SI}\!
  E_a{} ^{\rm SI}\!H_b =\cdots +\sqrt{\ve_0\mu_0}
  \underbrace{{} ^{\rm SI}\!\a_*^{ab}}_{\rm dim.-less}{} ^{\rm SI}\!E_a
  {} ^{\rm SI}\!H_b 
  \nonumber\\& =& \cdots + \frac{1}{4\pi}\underbrace{ {}^{\rm G}\!
    {{\a}}^{ab}}_{\rm dim.-less}{}^{\rm G}\!
  E_a{} ^{\rm G}\!H_b\,.
\end{eqnarray}
In SI, $[ {}^{\rm SI}\!{\a}^{ab}]=s/m$, $[{}^{\rm SI}\!E_a]=V/m$, and
$[^{\rm SI}\!H_b]=A/m$. Then,
\begin{equation}
  [{}^{\rm SI}\!g] = \frac{VAs}{m^3} = \frac{J}{m^3} =
  \frac{kg}{m\, s^2} =10\, \frac{g}{cm\, s^2}=10 \,[^{\rm G}\!g]\,.
\end{equation}
On the other hand, in the Gaussian system, we have for the electric
field
\begin{equation}
[^{\rm G}\!E_a]=\frac{statvolt}{cm}=3\times 10^4\,\frac{V}{m}=3\times
10^4\,[^{\rm SI}E_a]
\end{equation}
and for the magnetic excitation
\begin{equation} [^{\rm
    G}\!H_a]=O\!e=\frac{1000}{4\pi}\frac{A}{m}=\frac{1000}{4\pi}\,[^{\rm
    SI}H_a]\,.
\end{equation}
Thus we have for the magnetoelectric moduli
\begin{equation}\label{transf}
[^{\rm G}\!\a^{ab}]=[^{\rm SI}\!\a^{ab}_*]=1=c\,[^{\rm
  SI}\!\a^{ab}]\approx 3\times 10^8\,\frac{m}{s}\,[^{\rm
  SI}\!\a^{ab}]\,.
\end{equation}
Accordingly, we have the rule that multiplying $^{\rm SI}\a^{ab}$,
given in $s/m$, by $c = 3 \times 10^8\, m/s$ yields the dimensionless
Gaussian value $^{\rm G}\a^{ab}$. Incidentally, in
some papers Heaviside-Lorentz (``rationalized Gaussian'') units are
still in use, see, e.g., Borovik-Romanov \& Grimmer \cite{A}, p. 139.
\begin{figure}\label{fig1}
\includegraphics[width=10cm]{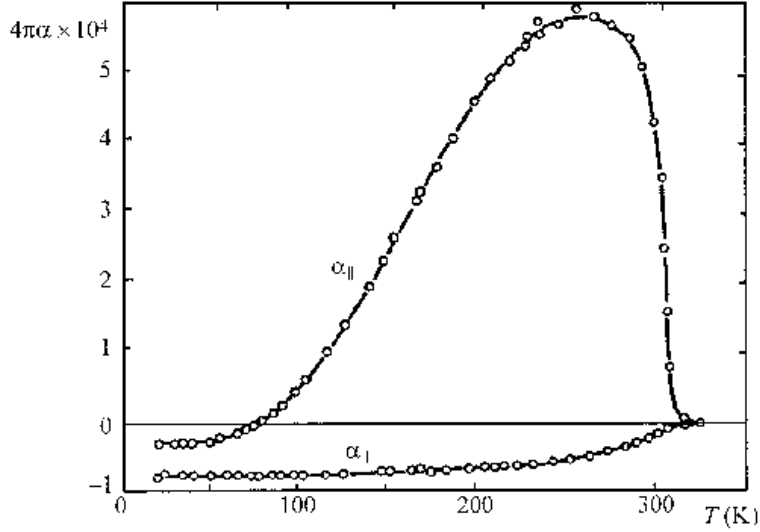}
\caption {The ME$_{\rm E}$ effect (linear magnetoelectric effect with
  electric field-induced magnetization) of Cr$_2$O$_3$: Temperature
  dependence of the magnetoelectric components $\a_{||}$ and $\a_\bot$
  according to Astrov \cite{Astrov}, see also the Tables
  \cite{Tables}.}
\end{figure}

\subsection{Astrov, Rado \& Folen, and Wiegelmann et al.}\label{sec42}

In our task to determine the pseudoscalar $\widetilde{\a}$, we can
take recourse to already published experimental data. Our main sources
are Astrov \cite{Astrov} (Fig.1) for the {\it electrically} induced
magnetoelectric effect (called ME$_{\rm E}$ in future) and Rado \&
Folen \cite{RadoFolen} for the {\it magnetically} induced
magnetoelectric effect (ME$_{\rm H}$), see also O'Dell \cite{O'Dell}
and \cite{Astrov0,AstrovErmakov,Date,Folen,Rado}. In both
investigations single crystals of Cr$_2$O$_3$ were used. In the
ME$_{\rm E}$ experiments \cite{Astrov,RadoFolen}, Eqs.(\ref{BE1}) to
(\ref{BE3}) were verified ($H$ switched off) and in the ME$_{\rm H}$
experiments \cite{RadoFolen} Eqs.(\ref{DH1}) to (\ref{DH3}) ($E$
switched off). In particular, Rado \& Folen made both type of
experiments and found that the magnetoelectric moduli $\a_\bot$ and
$\a_{||}$ for ME$_{\rm E}$ experiments coincide with those of the
ME$_{\rm H}$ experiments. This proves the vanishing of the skewon part
of the constitutive tensor $\chi^{\lambda\nu\sigma\kappa}$ for
Cr$_2$O$_3$.

Accordingly, these experiments confirmed Dzyaloshinskii's theory for
Cr$_2$O$_3$ below the spin-flop phase. Further experiments were then
done mainly for the ME$_{\rm H}$ case since (i) it is easier to
conduct an experiment with very high magnetic fields rather than with
high electric fields. (ii) Even at low $\mu_0H$ fields, say below 1
{\em tesla}, the ME$_{\rm H}$ effect needs no calibration of the
measuring system, contrary to the ME$_{\rm E}$ case!  With the
quasi-static method, as $D_a = - \partial g/\partial E_a =\a^{ab}H_b$
and $D_a = Q/S_a ,\,\a^{ab} = Q/(S_a H_b)$. The charges Q, usually in
the $pC$ range, are measured with a high input impedance electrometer,
the magnetic field $\mu_0H_b$ with a Hall probe and the area $S_a$ of one
of the electrode by taking a picture of it. (iii)
 Quasi-static as well
as AC measurements can be done easily. For ME$_{\rm E}$ quasi-static
experiments, a SQUID must be used, see Kita \cite{Kita2,Kita}.

\begin{figure}\label{fig2}
\includegraphics[width=12cm,height=9cm]{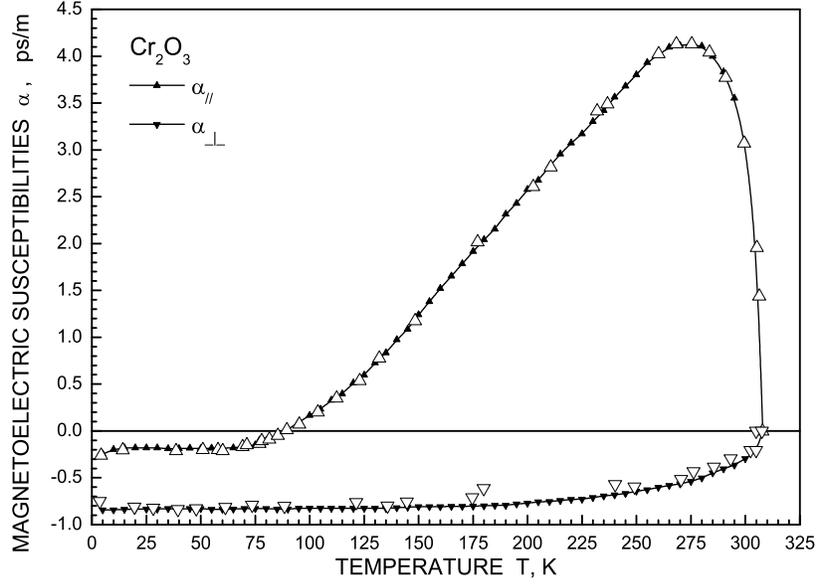}
\caption{The ME$_{\rm H}$ effect (linear magnetoelectric effect with
  magnetic field-induced polarization) of Cr$_2$O$_3$: Plot of
  $\a_{||}(T)$ and $\a_\bot(T)$ after digitalization and interpolation
  (small full triangles) of Fig.2.8, p.41, of Wiegelmann
  \cite{WiegelmannDr}, (see also Fig.2, p.143, Wiegelmann et al.\
  \cite{Wiegelmann}). The sign of $\a_\bot(T)$ was set negative
  according to Astrov \cite{Astrov}.  The curves (B-Splines) are only
  guides for the eyes.}
 \end{figure}

 With either methods, ME$_{\rm H}$ or ME$_{\rm E}$, (i) best results
 are obtained with gold electrodes rather than with silver paste ones.
 (ii) The resistivity of the sample must be high enough, especially
 when the ME$_{\rm H}$ quasi-static method of measurements is used.
 (iii) One has to make sure that the antiferromagnetic magnetoelectric
 crystal forms a single domain by cooling it in appropriate
 simultaneous electric and magnetic fields through the N\'eel
 temperature; this is the so-called magnetoelectric annealing, see
 \cite{Martin1}, \cite{Martin2}, and \cite{O'Dell}, p.124.
 Magnetoelectric crystals may also be (weakly) ferromagnetic or
 ferrimagnetic.

Our third main source of information are the measurements of Wiegelmann
et al.\ \cite{Wiegelmann}, see also \cite{WiegelmannDr,Wiegelmann2}.
He took magnetic fields $B$ as high as 20 {\it tesla} and measured
from liquid Helium up to room temperature. Wiegelmann et al.\ took a
quasi-static magnetic field and thereby disproved explicitly claims by
Lakhtakia \cite{Akhlesh1} that measurements with magnetic fields of
some kilo hertz cannot be extrapolated to static measurements. The
$\a_{zz}$ values of Wiegelmann et al.\ \cite{Wiegelmann} were in very
good agreement with independent $\a_{zz}$ measurement presented below
(Sec.\ref{sec43}).  However, the sign of $\a_\bot(T)$ relative to
$\a_{||}(T)$ was left open. Hence we took that from Astrov
\cite{Astrov}.

The values of $\a_\bot(T)$ and $\a_{||}(T)$ of Fig.2 are thus taken
from Wiegelmann \cite{WiegelmannDr}, Fig.2.8, p.41, (see also
Wiegelmann et al.\ \cite{Wiegelmann} (Fig.2, p.143)) after
digitalization, interpolation, and correction for the relative sign.
These values are given here in SI units in $ps/m$ (picosecond/meter).

Independently, see Sec.\ref{sec4}, Rivera measured quasistatically at
133 $H\!z$ $\a_{||}$ of Cr$_2$O$_3$ between $1.6\, K$ and $305\, K$.
He normalized that at $T = 275\, K$, the temperature of the maximum
value of $\a_{||}$.  He found the maximum value of $ \a_{||}\mbox{(at
  $275\, K$)} = 4.13\; ps/m$. As we saw in (\ref{transf}), $^{\rm
  SI}\a^{ab}$, given in $s/m$ must be multiplied by the speed of
light\footnote{The speed of light in SI is $c\approx 2.99792\times
  10^8\;m/s$.  Furthermore, in SI, $\mu_0=4\pi\times 10^{-7}\;\Omega
  s/m$.} in order to yield the dimensionless $^{\rm
  SI}\a_*^{ab}$. Dropping again the star, we have
\begin{equation}\label{value2}
  \a_{||}\,({\rm at}\, 275\, K) = 4.13\times 10^{-12}\,\frac{s}{m}\times
  2.99792\times 10^8\,\frac{m}{s}\,\approx 1.238\times 10^{-3}\,.
\end{equation}

\begin{figure}\label{fig3}
\includegraphics[width=7cm,height=7cm]{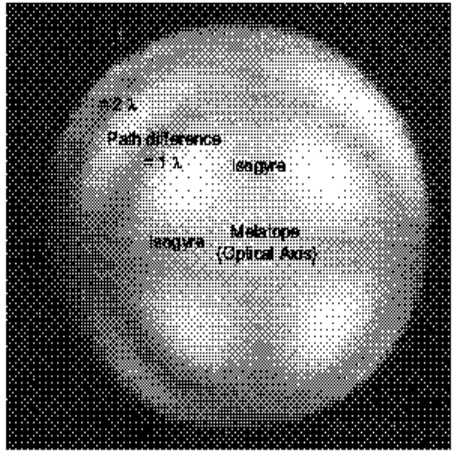}
\caption{Conoscopic pictures, in the near infrared, of a Cr$_2$O$_3$
  platelet at room temperature and between linear crossed polarizers,
  prepared for measuring the $\a_{||}(T)$ coefficient. We can see one
  or two isochromatic curve(s), the circle(s), the cut being slightly
  inclined, less than $3^\circ$ relatively to the optical axis, the
  $z$ axis.  At the center of the cross, the melatope emerges as the
  optical axis with a possible rotation of the $\mathbf{E}$ vector of
  the light along this axis, evidenced by a lighter center (Zeiss
  objective $125\times/1.30$ oil P, and condenser with head and oil,
  numerical aperture =  $1.30$).}
\end{figure}

\begin{figure}\label{fig4}
\includegraphics[width=10cm,height=7.5cm]{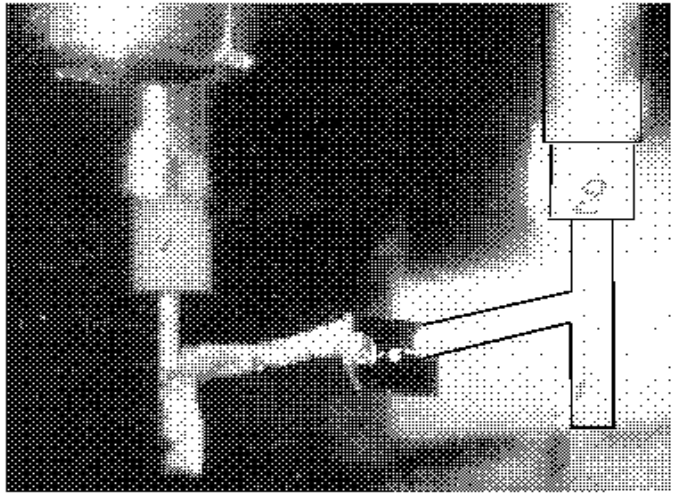}
\includegraphics[width=10cm,height=7.5cm]{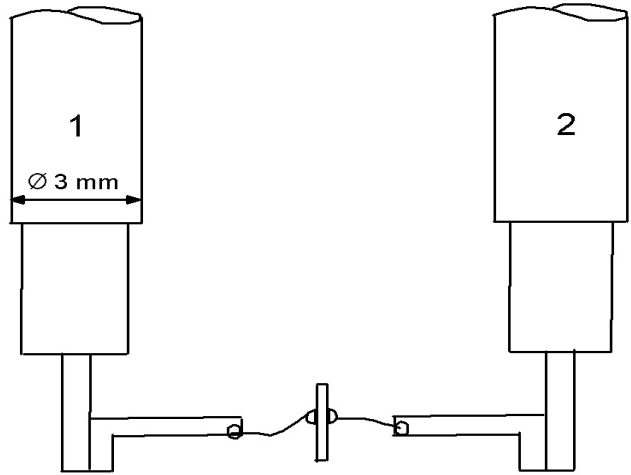}
\caption{a) A Cr$_2$O$_3$ platelet ($S_z = 4.70\, mm^2, th_z = 260
  \,\mu m$) for $\a_{||}(T)$ measurements, connected to two low noise
  coaxial cables \#1 and \#2. One can see, by reflection on the
  platelet, the gold wire and the Cu wire \#2 to the right. Black
  lines were added on the right of this unique Polaroid picture for
  clarity. b) A schematic drawing of the set-up depicted under a).}
\end{figure}

\begin{figure}\label{fig5}
\includegraphics[width=10.8cm,height=8.4cm]{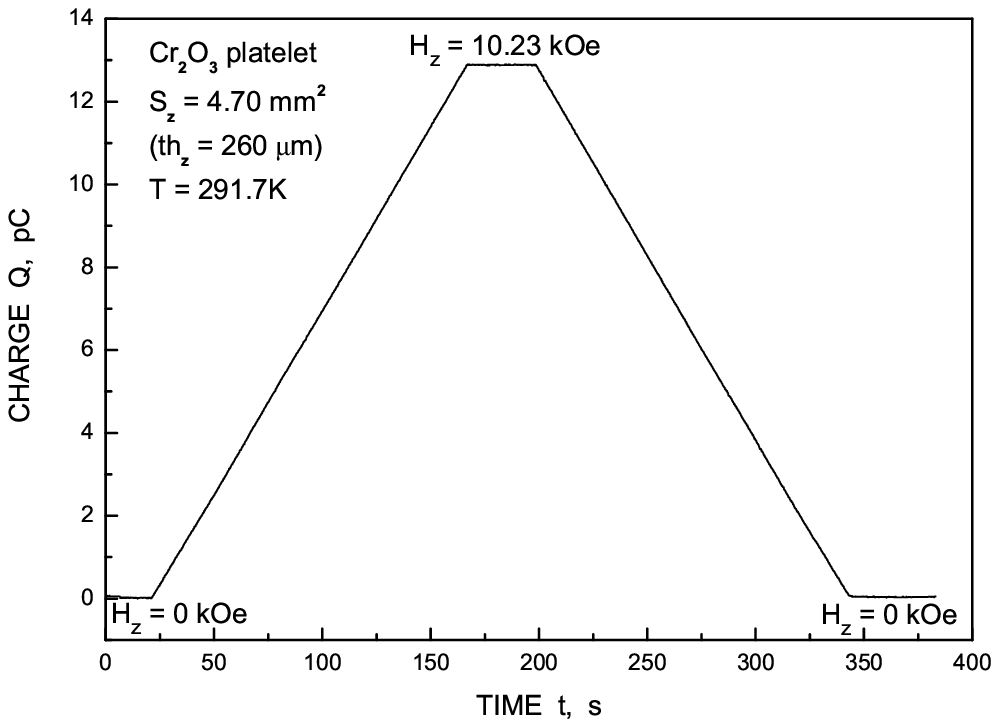}
\caption{Quasi-static ME$_{\rm H}$ experiment for Cr$_2$O$_3$, charge
  vs.\ time. The field $H$ varies linearly in time, from $0$ to
  $10\,kO\!e$ and from $10\,kO\!e$ to $0$, proving the linearity of
  the magnetoelectric effect in Cr$_2$O$_3$, here for $\a_{||}$(at 292
  $K$).}
\end{figure}

\subsection{Unpublished measurements of Rivera (1993)}\label{sec43}

One of us conducted quasi-static ME$_{\rm H}$ experiments on
Cr$_2$O$_3$ some years ago which were only briefly mentioned in
\cite{Rivera3}.  

After grinding and polishing a crystal, polarized light microscopy was
used, see, e.g., Wahlstrom \cite{Wahlstrom} or Hartshorne
\cite{Hartshorne}, to check the orientation of a perpendicular cut to
the optical axis.  Actually, conoscopy allowed such tests (Fig.3) in
the near infrared, $260\,\mu m$ thick Cr$_2$O$_3$ crystals being
absorbing in visible light. The symmetry was found to be uniaxial to
the accuracy of the conoscopic method, consistent with the symmetry
$\overline{3}{}'m'$ \cite{Brock}, although now it seems to be
$\overline{3}{}'$ \cite{DiMatteo}. The optical axis was computed to be
inclined less than $3^\circ$ away from the normal to the cut.
Incidentally, this platelet was then sent to Wiegelmann in Grenoble,
see \cite{Wiegelmann}.

For the ME$_{\rm H}$ experiments of $\a_{||}(T)$ presented below,
semi-transparent gold electrodes were evaporated on both sides of a
platelet with area (one side) $S_z = 4.70\, mm^2$ (thickness $th_z =
260\,\mu m$). On Fig.4, this Cr$_2$O$_3$ platelet is shown mounted
with its thin gold wires ({\O} $40\,\mu m$) on Cu wires and then on two
coaxial $50\,\Omega$ low noise cables \#1 and \#2, on a stainless
steel sample holder.  On the top left (forefront) of the picture, we
see the temperature sensor, a calibrated ($1.5\, K$ to $300\, K$)
Carbon Glass Resistor (CGR-2000, ``Lake Shore'', {\O} about $3\, mm$).
This sample holder was then inserted in a copper can with He exchange
gas, in an Helium bath cryostat.

As already mentioned, before the measurements, the crystal was always
cooled using the so-called magnetoelectric annealing with DC magnetic
and electric fields, see Martin, Anderson, and Schmid
\cite{Martin1,Martin2,Schmid189} in this general context. It was easy
to apply an electric field because the electrical resistivity of the
crystal was very high. We measured the charges $Q$ with a low noise
electrometer (Keithley, 642-LNFPA) and $H$ with a Hall probe. The $H$
was produced by an old 12-inch Varian ``V 4012-3B'' electromagnet with
Varian ``Mark I'' magnetic field regulator.  We measured at more than
130 temperature values.  Typically, the record of $Q$ was as follows
(quasi-static ME$_{\rm H}$ method): for about half a minute at zero
field, then increasing linearly $H$ with time, from $0$ to $10\,
kO\!e$ in about $2.5$ minutes, maintaining $H$ for half a minute at
maximum, decreasing $H$ linearly with time to $0$ and then maintaining
$H$ at $0$ for half a minute. This procedure was used to cancel any
shift of the base line, if any, see also Rivera
\cite{Rivera2,Rivera3}. The final curve looks like a volcano with a
flat top and straight sides, proving the linear character of the
ME$_{\rm H}$ effect for $\a_{||}$ of Cr$_2$O$_3$, see Fig.5 for T =
$291.7\, K$. Note the very good signal over signal + noise ratio.
Remember that $1\, O\!e = 1000/(4\pi)\, A/m \approx 79.6\, A/m$ and $H
= 10\, kO\!e$ yields $B = \mu_0 H = 1\, T$.  As $D_z = Q/S_z = \a_{||}
H_z$, we compute directly $\a_{||} = Q/(S_z H_z)$.

\begin{figure}\label{fig6}
\includegraphics[width=10cm,height=7.5cm]{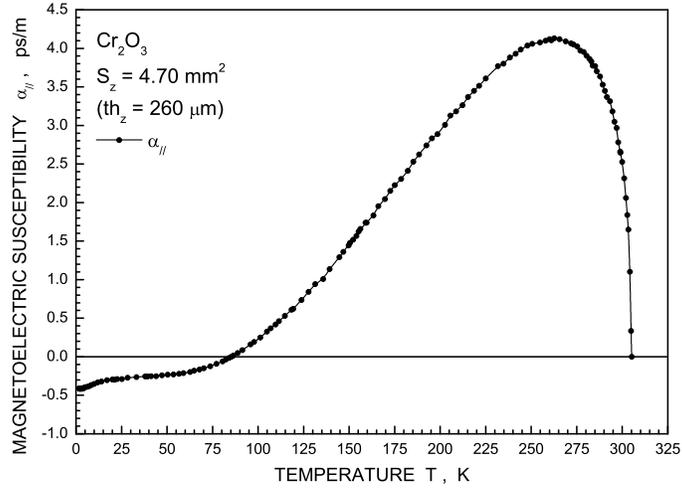}
\caption{Magnetoelectric susceptibility $\a_{||}(T)$ for
  Cr$_2$O$_3$, obtained by quasi-static ME$_{\rm H}$ experiments, $H$
  varying at each temperature from $0$ to $10\, kO\!e$ and back to
  $0$, as described in the text. The interpolation between the points
  was made by means of B-splines and is only a guide for the eyes.}
\end{figure}

In Fig.6, we show the curve of $\a_{||}(T)$), in SI units, i.e.,
$ps/m$. The curve, around $T = 10\, K$, has an elbow which could come
from a possible crystallographic and/or magnetic phase transition.
Further experiments should be done below $10\, K$ to clarify this
point.  Closely below $T_{\rm N}$, from $293\, K$ to $304\, K$, we
plotted (not shown) $\ln[\a_{||}(T)/\a_{||}^{\rm max}]$ vs.\
$\ln[(T_{\rm N}-T)/T_{\rm N}]$. It is a straight line, the slope gives
the exponent $\approx 0.34$ (thus about $1/3$). Astrov \cite{Astrov},
from the cut for measuring $\a_{\bot}(T)$ by a ME$_{\rm E}$
experiment, found an exponent $\approx 1/2$, as expected according to
the Landau theory.

In Table 1, Wiegelmann et al.\ \cite{Wiegelmann} compared the values
of the temperature $T$ at $\a_{||}=0$ (where the sign changes) and $T$
at $\a_{||}^{\rm max}$ that were obtained by different authors.  This
is better than giving an error on the results of $\a$ because with the
ME$_{\rm H}$ effect, the larger error, apart from the one on the area
of the crystal (about 3\% - 5\%), comes from the uncertainty about the
success in the magnetoelectric annealing, which could also be
influenced by the quality of the crystal.

\begin{figure}\label{fig7}
\includegraphics[width=10cm,height=7.5cm]{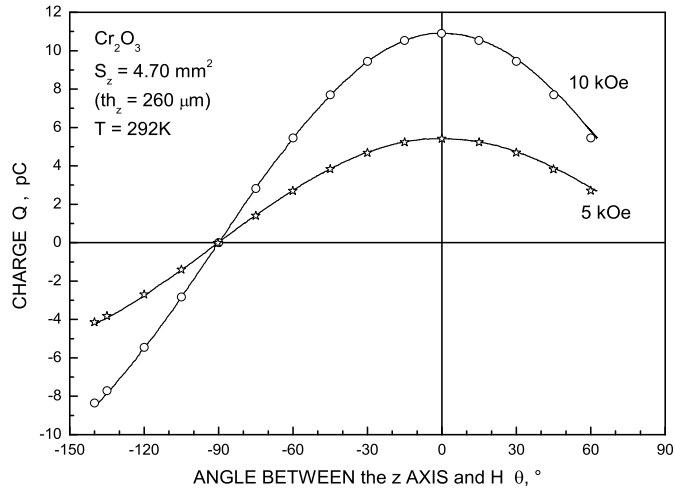}
\caption{ME$_{\rm H}$ experiments for Cr$_2$O$_3$ at $T = 292\,K$ and
  constant $H$ fields.  By rotating the electromagnet, the measured
  charges (continuous lines) follow cosine curves. Superimposed are
  normalized computed points (open circles at $H = 10\, kO\!e$, open
  stars at $H = 5\, kO\!e$). This proves that $\a_{zx}$ and $\a_{zy}$
  are very small or null.  At $\theta = 0^\circ$, the field $H$ is
  parallel to the optical axis (z axis).}
\end{figure}

In Fig.7, we display $Q(\theta)$ for $H = 5\, kO\!e$ and $10\,kO\!e$,
also at $T = 292\,K$, showing very good parts of cosine curves. The
angle $\theta$ is measured between the optical axis (the $z$ axis) and
the $H$ field direction. This demonstrates that $\a_{zx}$ and
$\a_{zy}$ are very small or even null, at least, at that temperature.
This again supports Dzyaloshinskii's theory, see (\ref{alpha}).  After
transformation from rectangular to polar coordinates, we could obtain
a continuous curve $\cos^2\theta$, similar to the one of Fig.3 given
by Astrov \cite{Astrov}\footnote{On p.731 of Astrov's paper
  \cite{Astrov} (English translation), 2nd column, the sentence
  ``Figure 3 shows \dots $103^\circ K$; the axis of rotation [i.e.,
  the $C_3$ axis] is perpendicular to the plane of the figure'' is
  wrong and `perpendicular' must be substituted by `parallel'.} for
the case $\a_{||}$\, at $T = 103\, K$.

\begin{figure}\label{fig8}
\includegraphics[width=10cm,height=7.5cm]{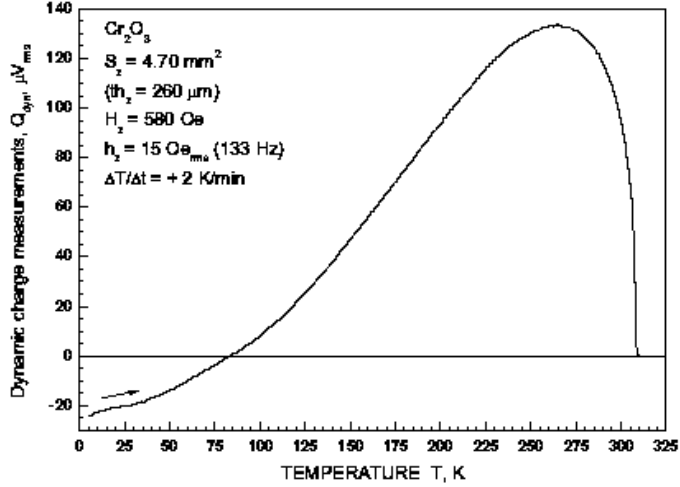}
\caption{Magnetoelectric susceptibility $\a_{||}(T)$ for
  Cr$_2$O$_3$ obtained by low frequency ($133\, Hz$) AC ME$_{\rm H}$
  experiments ($h_z\approx 18\, O\!e_{\rm rms}$).  Knowing $T\approx
  85.0\, K$ at $\a_{||}(T)=0$ from quasi-static experiments and the
  base line above $T_{\rm N}$, the curve was corrected for a small
  drift.}
\end{figure}

Fig.8 concerns the ME$_{\rm H}$ effect at a low frequency of $133\,
Hz$. It represents a dynamic measure of $\a_{||}(T)$, with an AC field
$h_z\approx 18\, O\!e_{\rm rms}$ (root mean square) superimposed to a
low DC field $H_z \approx 580\, O\!e$, just to maintain the domains
close to $T_{\rm N}$.  The rate of heating from $4.2\, K$ to $330\, K$
was $+ 2\, K/min$.  As shown in Fig. 4, the crystal was mounted
``floating'' between the coaxial cables \#1 and \#2.  As $Q = C U$, by
measuring with a ``lock-in'' amplifier (SR 530) the voltage induced by
the AC magnetic field $h_z$ we have a measure of the $Q(T)$ which is
proportional to $\a_{||}(T)$. To a first approximation, the
capacitance $C$ of the crystal was supposed to be independent of
temperature. If we superimposed the normalized curves at $T_{\rm max}$
of $\a_{||}(T)$, measured quasi-statically and dynamically, from $70\,
K$ to $270\, K$, the superposition is very good. The N\'eel
temperature found dynamically ($308\, K$) is $3\,K$ above the one
found quasi-statically ($305\, K$). Probably the heating rate ($+2\,
K/min$) was too high.

This concludes the presentation of our ME$_{\rm H}$ experiments on
Cr$_2$O$_3$ which are in very good agreement with the ones presented,
in particular, in \cite{Wiegelmann}.

\section{Extracting the pseudoscalar
  $\widetilde{\alpha}$}\label{sec5}

\subsection{The permeabilities $\mu_\bot$ and $\mu_{||}$}\label{sec51}

The magnetic susceptibility tensor of Cr$_2$O$_3$, namely
$\chi:=1-(1/\mu)\,$ or, in components,
$\chi_{ab}=\d_{ab}-\mu^{-1}_{ab}$, was determined by Foner
\cite{Foner} as a function of the temperature.\footnote{Don't mix up
  this 3-dimensional $\chi_{ab}$ with our 4-dimensional
  $\chi^{\lambda\nu\sigma\kappa}$ tensor.}  From the caption of his
Fig.8, we can take the static susceptibility perpendicular to the c
axis $\chi^{\rm g}_\bot$ of Cr$_2$O$_3$.  The superscript g stands for
specific (or mass) magnetic susceptibility.  Foner used the old unit
$emu/g$. From the inside of the front cover of Landolt-B\"ornstein
\cite{LandoltB} we learn that $1\, emu/g \equiv 1\, cm^3/g$.
Accordingly,
\begin{equation}\label{chi}
\chi^{\rm g}_\bot ({\rm at}\,4.2\, K) = 2.24
    \times 10^{-5}\,\frac{cm^3}{g}\,.
\end{equation}
Let us now determine the {\it volume} susceptibility $\chi^{\rm v}$.
Again from Landolt-B\"ornstein we take $^{\rm SI}\!\chi^{\rm
  v}=4\pi\,^{\rm G}\!\chi^{\rm v}$. The density for the mineral
eskolaite,\footnote{See, http://webmineral.com/data/Eskolaite.html .}
containing 94\% Cr$_2$O$_3$, is $5.23\,g/cm^3$ and for Cr$_2$O$_3$
ceramics\footnote{See,
  http://www.memsnet.org/material/chromiumoxidecr2o3bulk/ .} a bit
less, namely $5.21\,\,g/cm^3$. Then, in SI,
\begin{equation}\label{chi1}
  \chi^{\rm v}_\bot ({\rm at}\,4.2\, K)\,=\,
  {4\pi}\times2.24 \times
  10^{-5}\times 5.22\approx 1.47\times 10^{-3}\,.
\end{equation}
We can read off from Fig.8 of Foner \cite{Foner} that both, the
parallel and the perpendicular susceptibilities at the N\'eel
temperature $T_{\rm N}$ are about 13\% higher than $\chi^{\rm
  v}_\bot ({\rm at}\,4.2\, K)$. Consequently we find
\begin{eqnarray}\label{chiNeel}
  \chi^{\rm v}_\bot({\rm at}\;T_{\rm N})\approx\chi^{\rm v}_{||} 
  ({\rm at}\;T_{\rm N})\approx 1.13\times  \chi^{\rm v}_\bot
  ({\rm at}\;4.2\,K)\approx1.62\times 10^{-3}\,.
\end{eqnarray}
O'Dell \cite{O'Dell}, App.1, found the slightly higher value of
$\approx1.64\times 10^{-3}$. 

Now we can determine the permeabilities: Below and close to the N\'eel
temperature, we have
\begin{eqnarray}\label{perm1'}
  \mu_{\rm max}=\frac{1}{1-\chi^{\rm v}({\rm at}\;T_{\rm N})}
\approx 1+\chi^{\rm v}({\rm at}\;T_{\rm N})\approx 1.00162\,.
\end{eqnarray}
At 4.2 $K$, we find
\begin{eqnarray}\label{perm2'}
  \mu_{\bot}({\rm at}\,4.2 \,K)\approx 1+\chi^{\rm v}_\bot
  ({\rm at}\,4.2 \,K)\approx 1.00147\,.
\end{eqnarray}
Since $\chi_{||}$ is even smaller, we have $\mu\approx 1$. Because
Cr$_2$O$_3$ is antiferromagnetic, this was to be
expected.\footnote{\;For boracites, being antiferromagnetics with weak
  ferromagnetism, the assumption $\mu\approx 1$ is certainly not as
  good as for Cr$_2$O$_3$. But even with a factor of ten (say
  $\mu\approx 1.02$ instead of $1.0016$), the error of the areas on
  the small boracite crystals are certainly greater than that on
  $\mu$.}

\begin{figure}\label{fig9}
\includegraphics[width=10cm,height=7.5cm]{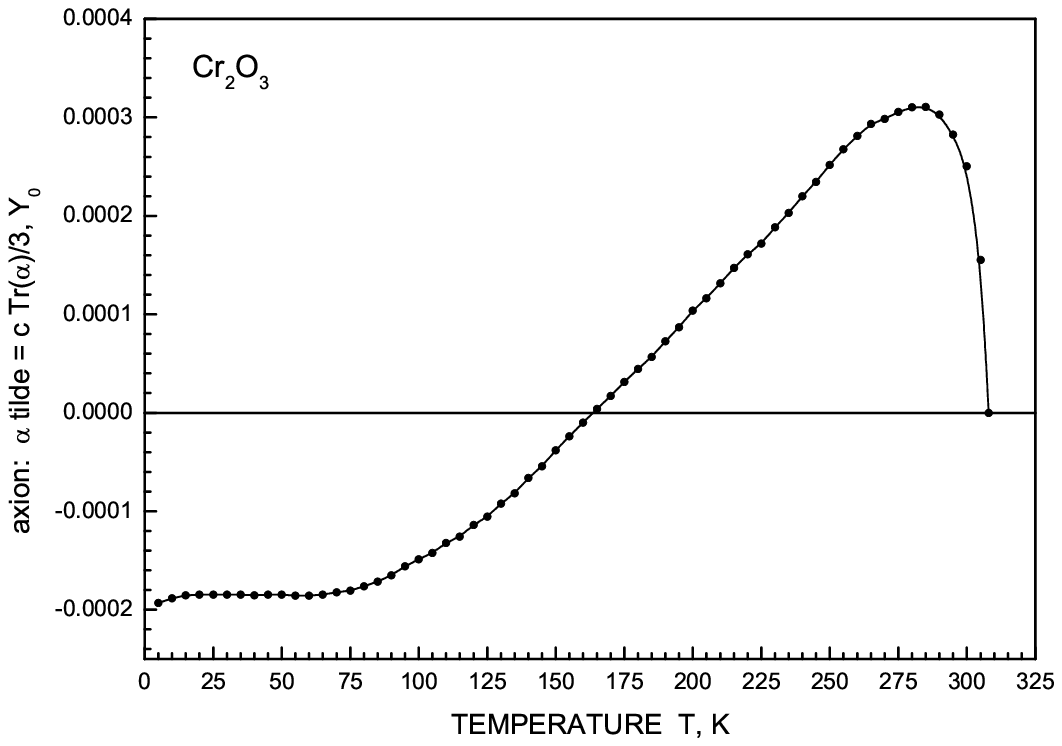}
\caption{The pseudoscalar or axion piece $\widetilde{\a}$, see
  Eq.(\ref{pss2}), of the constitutive tensor
  $\chi^{\lambda\nu\sigma\kappa}$ of Cr$_2$O$_3$ in units of
  $Y_0=1/Z_0$ as a function of the temperature $T$ in {\it kelvin};
  here $Z_0$ is the vacuum impedance which, in SI, is $\approx 377$
  {\it ohm}.}
\end{figure}

\subsection{The pseudoscalar (or axion piece) $\widetilde{\a}$}

Now we can come back to (\ref{axion}). Since
$\mu_\bot\approx\mu_{||}\approx 1$, the pseudoscalar (or axion piece)
of the constitutive tensor $\chi^{\lambda\nu\sigma\kappa}$ becomes
$\widetilde{\a}\,\widetilde{\epsilon}^{\lambda\nu\sigma\kappa}$, with 
\begin{equation}\label{result}
\widetilde{\a} \approx \frac 13\left(2\,{\a_\bot}+ 
      {\a_{||}} \right)\sq2\,. 
\end{equation}
It is the arithmetic mean of the trace of the magnetoelectric tensor
$^{\rm rel}\a^a{}_b$ of (\ref{alpha}). Going back to Fig.2, we can
take the values of $\a_\bot$ and $\a_{||}$ and compute
$\widetilde{\a}$. The result is plotted in Fig.9. As we can
see, for temperatures of up to about 163 $K$, the pseudoscalar is
negative, for higher temperatures positive until it vanishes at the
N\'eel temperature of about 308 $K$.  For example, at 285 $K$, we find
$\widetilde{\a}_{{\rm max}}\approx 1.035\,{ps/m}$.
Multiplied by $c$ we get
\begin{eqnarray}\label{result1}
  \widetilde{\a}_{{\rm max}}\;\mbox{(at 285 $K$)}&\approx& 3.10\times 
  10^{-4}\;\sq2\approx 
  \frac{1}{3226\;Z_0}\nonumber\\
&\stackrel{\rm SI}{\approx}& 8.22\times 
  10^{-7}\;\frac{1}{\Omega}\,{\approx}\,\frac{1}{1.216\,M\Omega}\,.
\end{eqnarray}

\section{Pseudoscalar or axion piece of the magnetoelectric
  susceptibility of Cr$_2$O$_3$ violates $P$ and $T$, it carries no
  electromagnetic energy density}\label{sec6}

As has been pointed out by Janner \cite{Janner2}, amongst others, the
violation of the invariances under space reflection (parity $P$) and
time inversion ($T$) of the corresponding crystal, in our case
Cr$_2$O$_3$, are necessary conditions for the emergence of the
magnetoelectric effect. The same is true for the emergence of the
pseudoscalar or axion piece $\widetilde{\a}$. The constitutive
relation for the axion piece alone can be read off from
(\ref{explicit3}) and (\ref{explicit4}) as
\begin{eqnarray}\label{Ax1}
D^a&=&+\widetilde{\a}\,B^a\,,\\
H_a&=&-\widetilde{\a}\,E_a\,,\label{Ax2}
\end{eqnarray}
see \cite{Birkbook}, Eqs.(D.1.112) and (D.1.111). If we denote, as in
crystallography, a space reflection by $\overline{1}$ and a time
inversion by $1'$, see Janner \cite{Janner2}, then we have
\begin{eqnarray}\label{refl1}
&&\overline{1}D^a=-D^a\,,\;\overline{1}H_a=H_a\,,\;
\overline{1}E_a=-E_a\,,\;\overline{1}B^a=B^a\,,\\
&&1'D^a=D^a\,,\;1'H_a=-H_a\,,\;1'E_a=E_a\,,\;1'B^a=-B^a\,,
\end{eqnarray}
see also Marmo et al.\ \cite{Bepe}. If we now apply a space reflection
to (\ref{Ax1}) and (\ref{Ax2}), then they transform into their
negatives,
\begin{eqnarray}\label{Ax1'}
D^a&=&-\widetilde{\a}\,B^a\,,\\
H_a&=&+\widetilde{\a}\,E_a\,.\label{Ax2'}
\end{eqnarray}
The same is true for a time inversion. In other words, the
constitutive relations for the axion piece {\em violate $P$ and $T$
  invariance,} see also \cite{Dz2,Toledano}. They are only invariant
under the combined $PT$ transformation. Consequently, the violation of
$P$ and $T$ invariance is an essential characteristics of the axion
piece $\widetilde{\a}$.

The first, to our knowledge, who tried to utilize the constitutive
laws (\ref{Ax1}) and (\ref{Ax2}) for vacuum electrodynamics was
Schr\"odinger \cite{Schrodinger}, p.25. He made the ansatz, compare
(\ref{constit7}),
\begin{equation}
  \mathfrak{G}^{\lambda\nu}=\frac{1}{2}\,\widetilde{\epsilon}
  ^{\lambda\nu\sigma\kappa}\,F_{\sigma\kappa}\quad\mbox{or}\quad
  \widetilde{G}_{\mu\nu}=F_{\mu\nu}\,.
\end{equation} 
A look at (\ref{Gtwiddle}) and (\ref{F}) shows that this, for
$\widetilde{\a}=1$, yields the laws (\ref{Ax1}) and (\ref{Ax2}).  But
Schr\"odinger rejected it as being unphysical if taken for vacuum
electrodynamics.

Another property is characteristic for the axion piece: It doesn't
contribute to the electromagnetic energy density. This can be seen
easily since the energy density in electrodynamics is $\frac
12\,(D^aE_a+H_aB^a)$. If (\ref{Ax1}) and (\ref{Ax2}) are substituted,
this expression vanishes. But even more so, also the energy flux
density vanishes. In order to prove this, we turn to the
energy-momentum tensor $\mathfrak{T}_\lambda{}^\nu$ of the electromagnetic
field that is built up from the energy density $\mathfrak{T}_0{}^0$, the
energy flux density $\mathfrak{T}_0{}^b$ (with $b=1,2,3$), the momentum
density $\mathfrak{T}_a{}^0$ (with $a=1,2,3$), and the momentum flux
density $\mathfrak{T}_a{}^b$ according to
\begin{equation}\label{em0}
\mathfrak{T}_\lambda{}^\nu=\begin{pmatrix}
\text{energy d.}& \text{energy flux d.}\\
\text{momentum d.} &\text{momentum flux d.}
\end{pmatrix}=\begin{pmatrix}
\mathfrak{T}_0{}^0 & \mathfrak{T}_0{}^b\\
\mathfrak{T}_a{}^0 &\mathfrak{T}_a{}^b
\end{pmatrix}\,.
\end{equation}
$\mathfrak{T}_0{}^b$ is also called the Poynting flux and $\mathfrak{T}_a{}^b$
the Maxwell stress. The energy-momentum tensor reads [cf.\ Post
\cite{Post}, eq.(9.55)]
\begin{equation}\label{em1}
\mathfrak{T}_\lambda{}^\nu=\mathfrak{L}\d^\nu_\lambda
-\mathfrak{G}^{\nu\sigma}F_{\lambda\sigma}\,,
\end{equation}
with
\begin{equation}\label{Lagrangian1}
\mathfrak{L}:=\frac 14\,\mathfrak{G}^{\sigma\tau}F_{\sigma\tau} =\frac
12({H_a}{B^a}-{D^a}{E_a})\,;
\end{equation}
the last equation can be read off directly from (\ref{frakG}) and
(\ref{F}). With some algebra, (\ref{em1}) can be rewritten as
\begin{equation}\label{em2}
\mathfrak{T}_\lambda{}^\nu=\frac
14\,\widetilde{\epsilon}^{\>\nu\mu\rho\sigma}(
\widetilde{G}_{\lambda\mu}F_{\rho\sigma}-F_{\lambda\mu}
\widetilde{G}_{\rho\sigma})\,,
\end{equation}
see \cite{Birkbook}, eq.(B.5.40). Since the laws (\ref{Ax1}) and
(\ref{Ax2}) can be put together in the manifestly covariant form 
\begin{equation}\label{Ax3}
\widetilde{G}_{\mu\nu}=\widetilde{\a}\,F_{\mu\nu}\,,
\end{equation}
we see immediately from (\ref{em2}) that 
\begin{equation}\label{em3}
\mathfrak{T}_\lambda{}^\nu\mbox{(of axion piece $\widetilde{\a}$)} =0\,.
\end{equation}
Thus, in particular, the electromagnetic energy density
$\mathfrak{T}_0{}^0$ of the axion piece vanishes.

\section{Other substances and symmetries permitting
  magnetoelectricity with the axion piece}\label{sec7}

In the present article the relativistic analysis is based on data of
the antiferromagnet Cr$_2$O$_3$, because it represents so far probably
the best studied magnetoelectric material and has diagonal components
of the linear magnetoelectric effect tensor $\a$, see (\ref{alpha}).
However, other materials and symmetries could have served the same
purpose, in principle. Among the 122 Heesch-Shubnikov point groups 58
ones are permitting the linear magnetoelectric effect
\cite{Schmid1974}, and therefrom 32 ones possess diagonal components
of the magnetoelectric tensor $\a$ \cite{A,Rivera3,Schmid203}.
Strictly speaking, our magnetoelectric tensor $^{\rm rel}\a$ in
(\ref{alpha}) belongs to the $EB$ scheme, see (\ref{explicit3}) and
(\ref{explicit4}), whereas the corresponding tensor in the literature
\cite{B} is the one of the $EH$ scheme. However, as we can see in
(\ref{alpha}), because of $\mu\approx 1$ the differences are marginal
and don't touch our arguments.

One can distinguish three types of diagonals (for the complete set of
magnetoelectric tensors see, e.g., refs.\ \cite{Rivera3},
\cite{Schmid203}, and \cite{A}, Table1.5.8.1, for the examples cited,
see ref.\ \cite{A}, Table 1.5.8.2, except for \cite{Goulon} and
\cite{B}):
\begin{enumerate}
\item\hspace{-8pt}) 19 point groups with:
  $\alpha_{11}\neq\alpha_{22}\neq\alpha_{33}$

        Examples:

        Point group $m'm'm'$ : DyAlO$_3$, GdAlO$_3$, TbAlO$_3$

        Point group $m'$ : Ni$_3$B$_7$O$_{13}$I

        Point group $m'm'2$ : Cu$_3$B$_7$O$_{13}$Cl

\item\hspace{-8pt}) {8 point groups with:
        $\alpha_{11}=\alpha_{22}\neq\alpha_{33}\;$ ($\alpha_{11} =
        \alpha_{22} = \alpha_\bot$, $\alpha_{33} = \alpha_{||}$)

        Examples:

        Point group $\overline{3}{}'m'$:
        [Cr$_2$O$_3$,]\footnote{\,Whereas neutron diffraction, as
          determined by Brockhouse \cite{Brock} and listed in
          \cite{A}, can only ``see'' the higher symmetry point group
          $\overline{3}{}'m'$, the detection of X-ray {\it
            magnetochiral dichroism} in Cr$_2$O$_3$
          \cite{Goulon,DiMatteo} requires the assumption of the lower
          symmetry point group $\overline{3}{}'$, thus superseding
          Brookhouse's earlier result. In contrast to the point groups
          $\overline{3}{}'m'$ and $\overline{3}{}'$, the monoclinic
          point group $2'/m$ \cite{X,Y} of the magnetic field-induced
        spin-flop phase of Cr$_2$O$_3$ does not allow diagonal
        magnetoelectric coefficients, but only off-diagonal ones,
        consistent with the magnetoelectric detection of a spontaneous
        toroidal moment \cite{Z}.}\, Nb$_2$Mn$_4$O$_9$,
      Nb$_2$Co$_4$O$_9$, Ta$_2$Mn$_4$O$_9$, Ta$_2$Co$_4$O$_9$

        Point group $\overline{3}{}'$: Cr$_2$O$_3\>$\,\cite{Goulon}

        Point group $4/m'm'm'$: Fe$_2$TeO$_6$}

\item\hspace{-8pt}) 5 point groups with:
        $\alpha_{11}=\alpha_{22}=\alpha_{33}$

        Examples:

        Point group $\overline{4}{}'3m'$ (expected): 
Cr$_3$B$_7$O$_{13}$Br, Cr$_3$B$_7$O$_{13}$I\;\,\cite{B}
\end{enumerate}

Thus, it is clear that the pseudoscalar $\widetilde{\a}$ occurs in
quite a number of different substances. Its existence can no longer be
denied.

\section{Discussion}\label{sec8}

The structure of the constitutive law (\ref{Ax3}) is not
unprecedented, as we already discussed in \cite{Postconstraint}. In
electrical engineering, in the theory linear networks, more
specifically in the theory of two ports (or four poles), Tellegen
\cite{Tellegen1948,Tellegen1956/7} came up with the new structure of a
{\it gyrator}, which is defined via
\begin{eqnarray}\label{gyrator}
&&v_1=-s\,i_2\,,\nonumber\\
&&v_2=\hspace{10pt}s\,i_1\,,
\end{eqnarray}
where $v$ are voltages and $i$ currents of the ports 1 and 2,
respectively. Let us quote from Tellegen \cite{Tellegen1956/7}, p.189:
``The ideal gyrator has the property of `gyrating' a current into a
voltage, and vice versa.  The coefficient $s$, which has the dimension
of a resistance, we call the gyration resistance; $1/s$ we call the
gyration conductance.'' The gyrator is a nonreciprocal network element.

If we turn to the electromagnetic field, then because of dimensional
reasons the quantities related to the {\it currents} $i_1,\,i_2$ are
the excitations $D^a,\,H_a$ and the quantities related to the {\it
  voltages} $v_1,\,v_2$ the field strengths $E_a,\,B^a$.  Then we find
without problems straightforwardly the relations
\begin{eqnarray}\label{gyrator*}
&&E_a=-s\,H_a\,,\nonumber\\
&&B^a=\hspace{10pt}s\,D^a\,.
\end{eqnarray}
If we rename the admittance $s$ according to $s=1/\widetilde{\a}$, then
(\ref{gyrator}) and (\ref{Ax1}),(\ref{Ax2}) coincide. Without the
least doubt, the gyrator is in the theory of two ports what the axion
piece is in magnetoelectricity. The axion piece `rotates' the
excitations, modulo an admittance, into the field strengths, as the
gyrator the currents into voltages.

These analogies or rather isomorphisms carry even further. In 2005,
Lindell \& Sihvola \cite{LindSihv2004a,LindSihv2004b}, see also
\cite{Ismobook}, introduced the new concept of a {\it perfect
  electromagnetic conductor} (PEMC). It also obeys the constitutive
law $\widetilde{G}_{\mu\nu}=\widetilde{\a}\,F_{\mu\nu}$ or
(\ref{gyrator*}). The PEMC is a generalization of the perfect electric
and the perfect magnetic conductor. In this sense, it is the `ideal'
electromagnetic conductor that can be hopefully built by means of a
suitable {\it metamaterial,} see Sihvola \cite{metaAri}. The
pseudoscalar $\widetilde{\a}$ is called Tellegen parameter by Lindell
et al., see \cite{Lindell1994}, p.13 (for a more general view, see
\cite{SihvolaLindell1995}); artificial Tellegen material has been
produced and positively tested by Tretyakov et al.\ \cite{Tretyakov},
amongst others.

Continuing with our search for isomorphisms, we turn to axion
electrodynamics, see Ni \cite{Ni}, Wilczek \cite{Wilczek87}, and, for
more recent work, Itin \cite{Itin2004,Itin2007}. If for vacuum
electrodynamics we add to the usual Maxwell-Lorentz expression
specified in (\ref{vacuum2}) an axion piece patterned after the last
term in (\ref{constit7}), then we have the constitutive law for axion
electrodynamics,\footnote{\;For $\a=$ const, the real part of Kiehn's
  {\it chiral vacuum} theory \cite{Kiehn2002} is a subcase of axion
  electrodynamics, see also \cite{Kiehn1,Kiehn2}.}
\begin{equation}\label{axel}
  \mathfrak{G}^{\lambda\nu}=\frac{1}{Z_0}\,\sqrt{-g}F^{\lambda\nu}+\frac
  12\,\widetilde{\a}\, \widetilde{\epsilon}^
  {\lambda\nu\sigma\kappa}F_{\sigma\kappa}\,.
\end{equation}
Alternatively, with the excitation pseudotensor (\ref{Gdual}) we find
\begin{equation}\label{vacuum3}
  \widetilde{G}_{\mu\nu}=\frac{1}{2Z_0}\,
  \widetilde{\epsilon}_{\mu\nu\kappa\lambda}\sqrt{-g}\,F^{\kappa\lambda}+
  \widetilde{\a}F_{\mu\nu}\,
\end{equation}
and, in exterior calculus,
\begin{equation}\label{axel*}
  \widetilde{G}=\left(\frac{1}{Z_0}\;^\star+\widetilde{\a}
  \right)F\,.
\end{equation} 
We discussed this `spacetime relation' and also the corresponding
Lagrangian in some detail in \cite{Birkbook} and
\cite{Postconstraint}.  The Hodge star operator $^\star$ is odd; it
transforms a form into a twisted form, and vice versa. Therefore we
could also denote it by $^{\widetilde{\star}}$ (we don't!). In
Cr$_2$O$_3$ we had $\widetilde{\a}\approx 10^{-4}/Z_0$. It is
everybody's guess what it could be for the physical vacuum. In
elementary particle theory one adds in the corresponding Lagrangian
also kinetic terms of the axion \`a la $\sim g^{\mu\nu}
\partial_\mu\widetilde{\a}\,\partial_\nu\widetilde{\a}$ and possibly a
massive term $\sim m_{\widetilde{\a}}\,\widetilde{\a}^2$.  However,
this hypothetical $P$ odd and $T$ odd particle has not been found so
far, in spite of considerable experimental efforts, see Davis et al.
\cite{Davis2007} and references given.

The axion shares its $P$ odd and $T$ odd properties with the
$\widetilde{\a}$ piece of Cr$_2$O$_3$, with the gyrator, and with the
PEMC. One may speculate whether an axion detector made of Cr$_2$O$_3$
crystals could enhance the probability of finding axions.

\subsection*{Acknowledgments} 
H.S.\ and J.-P.R.\ were supported (in the so far unpublished 1993
measurements) by the Swiss NSF. They had help from R.\ Boutellier and
E.\ Burkhardt (technical staff) which they gratefully acknowledge. The
Cr$_2$O$_3$ crystals were provided by M.\ Mercier, then at Grenoble.
One of us (F.W.H.) is very grateful for useful discussions with Yakov
Itin (Jerusalem), Ari Sihvola (Helsinki) and with M.~Braden,
T.~Nattermann and A.~Rosch (all from Cologne).  Financial support from
the DFG (HE 528/21-1) is gratefully acknowledged.


\begin{thebibliography}{99}

\bibitem{Ascher74} E.~Ascher, {\it Relativistic symmetries and
    lower bounds for the magnetoelectric susceptibility and the ratio
    of polarization to magnetization in a ferromagneto-electric
    crystal,} Physica Status Solidi {\bf B65} (1974) 677--688.

\bibitem{Astrov0} D.N.~Astrov, {\it The magnetoelectric effect in
    antiferromagnetics,} Sov.\ Phys.\ JETP {\bf 11} (1960) 708--709
  [Zh.\ Eksp.\ Teor.\ Fiz. {\bf 38} (1960) 984--985].

\bibitem{Astrov} D.N.~Astrov, {\it Magnetoelectric effect in chromium
    oxide,} Sov.\ Phys.\ JETP {\bf 13} (1961) 729--733 [Zh.\ Eksp.\
  Teor.\ Fiz. {\bf 40} (1961) 1035--1041].

\bibitem{AstrovErmakov} D.N.~Astrov and N.B.~Ermakov, {\it
    Quadrupole magnetic field of magnetoelectric Cr$_2$O$_3$,} JETP
  Lett. {\bf 59} (1994) 297--300.

\bibitem{Tables} A.~Authier, editor, {\it International Tables for
    Crystallography}, Vol.\ {\bf D}, Physical Properties of Crystals,
  Kluwer, Dordrecht/ Boston/London (2003).

\bibitem{Bergstrand} E.~Bergstrand, {\it Determination of the Velocity
    of Light}. In: {\sl Encyclopaedia of Physics,} S.~Fl\"ugge, ed.,
  Vol. {\bf XXIV}, Fundamentals of Optics, Springer, Berlin (1956),
  pp.1--43.

\bibitem{Born} M.~Born and E.~Wolf, {\it Priciples of Optics --}
  Electromagnetic theory of pro\-pagation, interference and diffraction
  of light, 7th (expanded) ed., Cambridge University Press, Cambridge,
  UK (1999).

\bibitem{A} A.S.~Borovik-Romanov and H.~Grimmer, {\it Magnetic
    Properties}, in \cite{Tables}, Sec.1.5, pp.105--149, see in particular,
  {\it Magnetoelectric effect}, Sec.1.5.8, pp.137--149.

\bibitem{Brock} B.N.~Brockhouse, {\it Antiferromagnetic structure
    in Cr$_2$O$_3$}, J.\ Chem.\ Phys. {\bf 21} (1953) 961--962.

\bibitem{Brown} L.~Brown, letter to the editor, Physics Today {\bf 54}
  (Jan.\ 2001) 13.

\bibitem{Date} M.~Date, J.~Kanamori, and M.~Tachiki, {\it Origin of
    magnetoelectric effect in Cr$_2$O$_3$,} J.\ Phys.\ Soc.\ Japan
  {\bf 16} (1961) 2589.

\bibitem{Davis2007} C.C.~Davis, J.~Harris, R.W.~Gammon,
  I.I.~Smolyaninov and K.~Cho, {\it Experimental Challenges Involved
    in Searches for Axion-Like Particles and Nonlinear Quantum
    Electrodynamic Effects by Sensitive Optical Techniques,}
  arxiv.org/abs/0704.0748 [hep-th].
 
\bibitem{deLange} O.L.~de Lange and R.E.~Raab, {\it Post's constraint
    for electromagnetic constitutive relations}, J.\ Opt.\ A: Pure
  Appl.\ Opt. {\bf 3} (2001) L23--L26.

\bibitem{DiMatteo} S.~Di Matteo and C.R.~Natoli, {\it Magnetochiral
    dichroism in Cr$_2$O$_3$}, Phys.\ Rev. {\bf B66} (2002) 212413 (4
  pages).

\bibitem{Dz1} I.E.~Dzyaloshinskii, {\it On the magneto-electrical
    effect in antiferromagnets,} J.\ Exptl.\ Theoret.\ Phys. (USSR)
  {\bf 37} (1959) 881--882 [English transl.: Sov.\ Phys.\ JETP {\bf
    10} (1960) 628--629].\medskip

\bibitem{Dz2} I.~Dzyaloshinskii, {\it Time parity violation in
    quantum and conventional models}, Ferroelectrics {\bf 161} (1994)
  253--255.

\bibitem{Fiebig2} M.~Fiebig, {\it Revival of the magnetoelectric
    effect}, J.\ of Phys. {\bf D38} (2005) R123--R152.

\bibitem{X} M.~Fiebig, D.~Fr\"ohlich, and H.J.~Thiele, {\it
    Determination of spin direction in the spin-flop phase of
    Cr$_2$O$_3$,} Phys.\ Rev. {\bf B54} (1996) R12681--R12684.

\bibitem{Flowers} J.L.~Flowers and B.W.~Petley, {\it Progress in our
    knowledge of the fundamental constants in physics\/,} {\sl Rep.\
    Progr.\ Phys.} {\bf 64} (2001) 1191-1246.

\bibitem{Folen} V.J.~Folen, G.T.~Rado, and E.W.~Stalder, {\it
    Anisotropy of the magnetoelectric effect in Cr$_2$O$_3$}, Phys.\
  Rev.\ Lett. {\bf 6} (1961) 607--608.

\bibitem{Foner} S.~Foner, {\it High-field antiferromagnetic resonance
    in Cr$_2$O$_3$}, Phys.\ Rev. {\bf 130} (1963) 183--197.

\bibitem{Gehring} G.A.~Gehring, {\it On the microscopic theory of the
    magnetoelectric effect,} Ferroelectrics {\bf 161} (1994) 275--285.

\bibitem{Goulon} J.~Goulon, A.~Rogalev, F.~Wilhelm, C.~Goulon-Ginet,
  P.~Carra, D.~Cabaret, and C.~Brouder, {\it X-ray magnetochiral
    dichroism: A new spectroscopic probe of parity nonconserving
    magnetic solids}, Phys.\ Rev.\ Lett.  {\bf 88} (2002) 237401
  (2002) (4 pages).

\bibitem{Hartshorne} N.H.~Hartshorne and A.~Stuart, {\em Practical
    Optical Crystallography,} Arnold, London (1971).

\bibitem{Zlatibor} F.W.~Hehl, Y.~Itin, and F.W.~Hehl, {\it Recent
    developments in premetric classical electrodynamics}, in: {\sl
    Proceedings of the 3rd} {\sl Summer School in Modern Mathematical
    Physics,} {\it 20-31 August 2004, Zlatibor, Serbia and
    Montenegro}, B.~Dragovich, Z.~Rakic and B.~Sazdovic, eds.,
  (Institute of Physics: Belgrade, 2005). In the series SFIN
  (Notebooks on Physical Sciences) XVIII: Conferences, A1 (2005)
  375-408; with updated references as arXiv.org/physics/0610221.

\bibitem{Birkbook} F.W.~Hehl and Yu.N.~Obukhov, {\it Foundations of
    Classical Electrodynamics --} Charge, flux, and metric
  (Birkh\"auser: Boston, MA, 2003).

\bibitem{Postconstraint} F.W.~Hehl and Yu.N.~Obukhov, {\it Linear
    media in classical electrodynamics and the Post constraint},
  Phys.\ Lett. {\bf A334} (2005) 249--259; arXiv.org/physics/0411038.

\bibitem{Rosenow} F.W.~Hehl, Yu.N.~Obukhov and B.~Rosenow, {\it Is the
    Quantum Hall Effect influenced by the gravitational field?} {\sl
    Phys.\ Rev.\ Lett.} {\bf 93} (2004) 096804 (4 pages);
  arXiv.org/cond-mat/0310281.

\bibitem{Itin2004} Y.~Itin, {\em Caroll-Field-Jackiw electrodynamics
    in the pre-metric framework,} Phys.\ Rev. {\bf D70} (2004) 025012
  (6 pages).

\bibitem{Itin2007} Y.~Itin, {\em Wave propagation in axion
    electrodynamics,} arXiv:0706.2991v1 [hep-th] (15 pages).

\bibitem{Jackson} J.D.~Jackson, {\it Classical Electrodynamics,} 3rd
  edition, Wiley, New York (1999).

\bibitem{Janner} A.~Janner, {\it Looking for a relativistic crystal
    optics,} Ferroelectrics {\bf 161} (1994) 191--206.

\bibitem{Janner2} A.~Janner, {\it On the relativistic symmetries of
    magnetoelectric quasi-moving crystals,} Physica {\bf B204} (1995)
  287--291.

\bibitem{Kamenetskii} E.O.~Kamenetskii, {\it Bianisotropics and
    electromagnetics}, arXiv.org/cond-mat/0601467.

\bibitem{Kiehn2002} R.M.~Kiehn, {\it The chiral vacuum}, preprint
  1997, updated 2002, see {\sl Cartan's Corner:} {\tt
    http://www22.pair.com/csdc/pdf/chiral.pdf} (11 pages).

\bibitem{Kiehn1} R.M.~Kiehn, {\it Chirality and helicity in terms of
    topological spin and topological torsion}, arXiv:physics/0101101
  (16 pages).

\bibitem{Kiehn2} R.M.~Kiehn, {\it A topological theory of the physical
    vacuum}, arXiv:gr-qc/0602118 (59 pages).
     
\bibitem{Kita} E.~Kita, {\it DC magnetoelectric effect measurements by
    a Squid magnetometer,} Ferroelectrics {\bf 162} (1994) 397--400.

\bibitem{Kita2} E.~Kita, A.~Tasaki and K.~Siratori, {\it Application
    of SQUID magnetometer to the measurement of magnetoelectric effect
    in Cr$_2$O$_3$}, Jpn.\ J.\ Appl.\ Phys. {\bf 18} (1979) 1361--1366.

\bibitem{Akhlesh1} A.~Lakhtakia, {\it On the genesis of the Post
    constraint in modern electromagnetism}, Optik {\bf 115} (2004)
  151--158; arXiv.org/physics/ 0403042.

\bibitem{Akhlesh2} A.~Lakhtakia, {\em Boundary-value problems and the
    validity of the Post constraint in modern electromagnetism,} Optik
  {\bf 117} (2006) 188--192.

\bibitem{LL} L.D.~Landau and E.M.~Lifshitz, {\it Electrodynamics
    of Continous Media}, Vol.8 of {\it Course of Theoretical
    Physics}, transl.\ from the Russian (Pergamon: Oxford, 1960).

\bibitem{LandoltB} Landolt-B\"ornstein, New Series, Vols. III/12b
  (1980) and III/12c (1982), Springer, Berlin.

\bibitem{Ismobook} I.V.~Lindell, {\it Differential
    Forms in Electromagnetics} (IEEE Press: Piscataway, NJ, and
  Wiley-Interscience, 2004).

\bibitem{LindSihv2004a} I.V.~Lindell and A.H.~Sihvola,
  {\it Perfect electromagnetic conductor}, J. Electromagn. Waves Appl.
  {\bf 19} (2005) 861--869.

\bibitem{LindSihv2004b} I.V.~Lindell and A.H.~Sihvola,
  {\it Transformation method for problems involving perfect
    electromagnetic conductor (PEMC) structures},
  IEEE Trans.\ Antennas Propag. {\bf 53} (2005) 3005--3011.

\bibitem{Lindell1994} I.V.~Lindell, A.H.~Sihvola, S.A.~Tretyakov,
  A.J.~Viitanen, {\it Electromagnetic Waves in Chiral and Bi-Isotropic
    Media}. Artech House, Boston (1994).

\bibitem{Bepe} G.~Marmo, E.~Parasecoli, and W.~Tulczyjew, {\it
    Space-time orientations and Maxwell's equations,} Rep. on Math.\ 
  Phys. (Toru\'n) {\bf 56} (2005) 209--248.  

\bibitem{Martin1} T.J.~Martin, {\em Antiferromagnetic domain switching
    in Cr$_2$O$_3$,} Phys.\ Lett. {\bf 17} (1965) 83--85.

\bibitem{Martin2} T.J.~Martin and J.C.~Anderson, {\em
    Antiferromagnetic domain switching in Cr$_2$O$_3$,} IEEE Trans.\
  Magn. {\bf 2} (1966) 446--449.

\bibitem{McGuire} T.R.~McGuire, E.J.~Scott, and F.H.~Grannis, {\it
    Antiferromagnetism in a Cr$_2$O$_3$ crystal}, Phys.\ Rev. {\bf 98}
  (1955) 1562.

\bibitem{NIST} National Institute of Standards and Technology (USA):
  {\tt http:// physics.nist.gov}\,.

\bibitem{Ni} W.-T.~Ni, {\it Equivalence principles and
    electromagnetism}, Phys.\ Rev.\ Lett. {\bf 38} (1977) 301--304.

\bibitem{measuringAxion} Yu.N.~Obukhov and F.W.~Hehl, {\it Measuring a
    piecewise constant axion field in classical electrodynamics},
  Phys.\ Lett. {\bf A341} (2005) 357--365; arXiv.org/physics/0504172.

\bibitem{Optik} Yu.N.~Obukhov and F.W.~Hehl, {\it On the
    boundary-value problems and the validity of the Post constraint in
    modern electromagnetism,} submitted to Optik (2007);
  arXiv.org/abs/0707.1112 [physics.class-ph].

\bibitem{O'Dell1966} T.H.~O'Dell, {\it Measurement of the
    magneto-electric susceptibility of polycrystalline chromium
    oxide}, Phil.\ Mag. {\bf 13} (1966) 921--933.

\bibitem{O'Dell} T.H.~O'Dell, {\it The Electrodynamics of
    Magneto-Electric Media,} North-Holland, Amsterdam (1970).

\bibitem{Panofsky} W.K.H.~Panofsky and M.~Phillips, {\it Classical
    Electricity and Magnetism}, 2nd ed., Addison-Wesley, Reading, MA
  (1962), and Dover, Mineola, New York (2005).

\bibitem{Y} R.V.~Pisarev, M.~Fiebig, and D.~Fröhlich, {\it Nonlinear
    optical spectroscopy of magnetoelectric and piezomagnetic
    crystals,} Ferroelectrics, {\bf 204} (1997) 1--21.

\bibitem{Z} Yu.F.~Popov, A.M.~Kadomtseva, D.V.~Belov, G.P.~Vorob'ev,
  and A.K. Zvezdin, {\it Magnetic-field-induced toroidal moment in the
    magnetoelectric Cr$_2$O$_3$,} JETP Letters {\bf 69} (1999)
  330-335.

\bibitem{Post} E.J.~Post, {\it Formal Structure of Electromagnetics}
    -- General Covariance and Electromagnetics (North Holland:
  Amsterdam, 1962, and Dover: Mineola, New York, 1997).

\bibitem{RaabSihvola1997} R.E.~Raab and A.H.~Sihvola, {\it On the
    existence of linear non-reciprocal bi-isotropic (NRBI) media,} J.\ 
  Phys.\ {\bf A30} (1997) 1335--1344.

\bibitem{Rado} G.T.~Rado, {\it Mechanism of the magnetoelectric
    effect in an antiferromagnet,} Phys.\ Rev.\ Lett. {\bf 6} (1961)
  609--610.

\bibitem{RadoFolen} G.T.~Rado and V.J.~Folen, {\it Observation of the
    magnetically induced magnetoelectric effect and evidence for
    antiferromagnetic domains}, Phys.\ Rev.\ Lett. {\bf 7} (1961)
  310--311.

\bibitem{RadoFolen62} G.T.~Rado and V.J.~Folen, {\it Magnetoelectric
    effects in antiferromagnetics,} J.\ Appl.\ Physics {\bf 33} (1962)
  1126--1132.

\bibitem{RaabBook} R.E.~Raab and O.L.~de~Lange, {\it Multipole Theory
    in Electromagnetism,} Classical, quantum, and symmetry aspects
  with applications, Clarendon Press, Oxford (2005).

\bibitem{Raith} W.~Raith, ed., {\em Bergmann-Schaefer, Lehrbuch der
    Experimentalphysik, Vol.2, Elektromagnetismus}, 9th rev.\ ed. (de
  Gruyter: Berlin, 2006).

 \bibitem{Rivera2} J.P.~Rivera, {\it Magnetoelectric effect in
     LiCoPO$_4$,} Ferroelectrics {\bf 161} (1994) 147--164.

\bibitem{Rivera3} J.P.~Rivera, {\it On definitions, units,
    measurements, tensor forms of the linear magnetoelectric effect
    and on a new dynamic method applied to Cr-Cl boracite},
  Ferroelectrics  {\bf 161} (1994) 165--180.

\bibitem{Schmid1974} H.~Schmid, {\it On a magnetoelectric
    classification of materials,} Int.\ J.\ Magnetism {\bf 4} (1973)
  337--361.

\bibitem{Schmid189} H.~Schmid, {\it On the possibility of
    ferromagnetic, antiferromagnetic, ferroelectric and ferroelastic
    domain reorientations in magnetic and electric fields},
  Ferroelectrics {\bf 221} (1999) 9--17.

\bibitem{Schmid203} H.~Schmid, {\it Magnetoelectric effects in
    insulating magnetic materials,} in: {\it Introduction to complex
    mediums for optics and electromagnetics}, W.S.~Weigl\-hofer and
  A.~Lakhtakia, eds,, pp.167--195, SPIE Press, Bellingham, WA, USA
  (2003).

\bibitem{B} H.~Schmid, {\it Some supplementing comments on the
    Proceedings of MEIPIC-5}, in: M.~Fiebig, V.V.~Eremenko and
  I.E.~Chupis (eds.), {\it Magnetoelectric Interaction Phenomena in
    Crystals}, Kluwer Academic Publishers, Dordrecht/Boston/London
  (2004) pp.1--34.

\bibitem{Schrodinger} E. Schr\"odinger, {\it Space-Time Structure}
  (Cambridge University Press: Cambridge, 1954).

\bibitem{Serdyukov} A.~Serdyukov, I.~Semchenko, S.~Tretyakov, and
  A.~Sihvola, {\em Electromagnetics of Bi-anisotropic Materials,}
    Theory and Applications, Gordon and Breach, Amsterdam (2001).

\bibitem{Ari1995} A.H.~Sihvola, {\it Are nonreciprocal bi-isotropic
    media forbidden indeed?}, IEEE Trans. Microwave Theory Techn. {\bf
    43} (1995) 2160--2162; see also the discussion on pp.2722--2724.

\bibitem{metaAri} A.H.~Sihvola, {\it Metamaterials in
    electromagnetics,} Metamaterials {\bf 1} (2007) 2--11.

\bibitem{SihvolaLindell1995} A.H.~Sihvola and I.V.~Lindell, {\it
    Material effects on bi-anisotropic electromagnetics,} IEICE
  Trans.\ Electron.\ (Tokyo) {\bf E78-C} (1995) 1383--1390.

\bibitem{SihTre} A.~Sihvola and S.~Tretyakov, {\em Comments on
    boundary problems and electromagnetic constitutive parameters,}
  Optik, 3 pages, to be published (2007).

\bibitem{Sokol} I.S.~Sokolnikoff, {\it Tensor Analysis,} Wiley, New
  York (1951).

\bibitem{Sommerfeld}A.~Sommerfeld, {\it Elektrodynamik.} Vorlesungen
  \"uber Theoretische Physik, Band 3 (Dieterich'sche
  Verlagsbuchhandlung: Wiesbaden, 1948). English translation: A.\
  Sommerfeld, {\it Electrodynamics,\/} Vol.\ 3 of Lectures in
  Theoretical Physics (Academic Press: New York, 1952).

\bibitem{Tamm} I.E.~Tamm, {\it Relativistic crystal optics and its
     relation to the geometry of a bi-quadratic form}, {\sl Zhurn.
     Ross. Fiz.-Khim. Ob.} {\bf 57}, n. 3-4 (1925) 209-224 (in
   Russian). Reprinted in: I.E. Tamm, {\it Collected Papers} (Nauka:
   Moscow, 1975) Vol.\ 1, pp.\ 33-61 (in Russian).  See also: I.E.
   Tamm, {\it Electrodynamics of an anisotropic medium in special
     relativity theory}, ibid, pp.\ 19-31; a short version therefrom
   appeared in German: L.~Mandelstam and J.~Tamm, Mathematische
   Annalen {\bf 95} (1925) 154--160.

 \bibitem{Tellegen1948} B.D.H.~Tellegen, {\it The gyrator, a new
    electric network element,} Philips Res.\ Rep. {\bf 3} (1948)
  81--101.

\bibitem{Tellegen1956/7} B.D.H.~Tellegen, {\it The gyrator, an
    electric network element,} Philips Technical Review {\bf 18}
  (1956/57) 120--124. Reprinted in H.B.G.\ Casimir and S.\ Gradstein
  (eds.) {\it An Anthology of Philips Research.} Philips'
  Gloeilampenfabrieken, Eindhoven (1966) pp.186--190.

\bibitem{Toledano} P.~Tol\'edano, {\it Magnetoelectric symmetry and
     the Landau theory of phase transitions,} Ferroelectrics {\bf 161}
   (1994) 257--273.

 \bibitem{Tretyakov} S.A.Tretyakov, S.I.Maslovski, I.S.~Nefedov,
   A.J.~Viitanen, P.A.~Belov, and A.~Sanmartin, {\em Artificial
     Tellegen particle,} Electromagnetics {\bf 23} (2003) 665--680.

 \bibitem{Wahlstrom} E.E.~Wahlstrom, {\it Optical Crystallography,}
   Wiley, New York (1979).

\bibitem{WiegelmannDr} H.~Wiegelmann, {\it Magnetoelectric effects in
    strong magnetic fields}, Ph.D.\ thesis, University of Konstanz
  (1994).

\bibitem{Wiegelmann2} H.~Wiegelmann, A.G.M.~Jansen, J.-P.~Rivera,
   H.~Schmid, A.A.~Stepanov, I.M.~Vitebsky, {\it Magnetoelectric
     effect of antiferromagnetic crystals in strong magnetic fields,}
   Physica {\bf B204} (1995) 292--297.

\bibitem{Wiegelmann} H.~Wiegelmann, A.G.M.~Jansen, P.~Wyder,
  J.-P.~Rivera, and H.~Schmid, {\it Magnetoelectric effect of
    Cr$_2$O$_3$ in strong static magnetic fields,} Ferroelectrics {\bf
    162} (1994) 141--146.

\bibitem{Wilczek87} F.\ Wilczek, {\it Two applications of axion
    electrodynamics}, {\sl Phys.\ Rev.\ Lett.} {\bf 58} (1987)
  1799-1802.

\end{thebibliography}
\end{document}